\input harvmac.tex 

\input epsf.tex

\newcount\figno
\figno=0
\def\fig#1#2#3{
\par\begingroup\parindent=0pt\leftskip=1cm\rightskip=1cm\parindent=0pt
\baselineskip=11pt
\global\advance\figno by 1
\midinsert
\epsfxsize=#3
\centerline{\epsfbox{#2}}
\vskip 12pt
{\bf Fig. \the\figno:} #1\par
\endinsert\endgroup\par
}
\def\figlabel#1{\xdef#1{\the\figno}}
\def\encadremath#1{\vbox{\hrule\hbox{\vrule\kern8pt\vbox{\kern8pt
\hbox{$\displaystyle #1$}\kern8pt}
\kern8pt\vrule}\hrule}}

%%%% Journals %%%%%%%%%%%%%%%%%%%%%%%%%%%%%%%%%%%%%%%%%%

\def\np#1#2#3{Nucl. Phys. {\bf B#1} (#2) #3}
\def\pl#1#2#3{Phys. Lett. {\bf #1B} (#2) #3}
\def\prl#1#2#3{Phys. Rev. Lett. {\bf #1} (#2) #3}
\def\cmp#1#2#3{Comm. Math. Phys. {\bf #1} (#2) #3}
\def\physrev#1#2#3{Phys. Rev. {\bf D#1} (#2) #3}
\def\ijmp#1#2#3{Int. J. Mod. Phys. {\bf A#1} (#2) #3}

\batchmode
  \font\bbbfont=msbm10
\errorstopmode
\newif\ifamsf\amsftrue
\ifx\bbbfont\nullfont
  \amsffalse
\fi
\ifamsf
\def\IR{\hbox{\bbbfont R}}
\def\IZ{\hbox{\bbbfont Z}}
\def\IF{\hbox{\bbbfont F}}
\def\IP{\hbox{\bbbfont P}}
\else
\def\IR{\relax{\rm I\kern-.18em R}}
\def\IZ{\relax\ifmmode\hbox{Z\kern-.4em Z}\else{Z\kern-.4em Z}\fi}
\def\IF{\relax{\rm I\kern-.18em F}}
\def\IP{\relax{\rm I\kern-.18em P}}
\fi
\def\pf{{\rm Pf ~}}
\def\ev#1{\langle#1\rangle}

\def\CW{{\cal W}}

\lref\nati{N. Seiberg, ``Exact Results on the Space of Vacua of Four
Dimensional SUSY Gauge Theories'', 
hep-th/9402044, \physrev{49}{1994}{6857}}%
\lref\sem{N. Seiberg, ``Electric-Magnetic Duality in Supersymmetric
Non-Abelian Gauge Theories'', hep-th/9411149 , \np{435}{1995}{129}.}%
\lref\ahw{I. Affleck, J. Harvey and E. Witten, ``Instantons and
(Super-)Symmetry Breaking in (2+1) Dimensions'', \np{206}{1982}{413}.}%
\lref\ads{I. Affleck, M. Dine and N. Seiberg, ``Dynamical
Supersymmetry Breaking in Supersymmetric QCD'', \np{241}{1984}{493}\semi
I. Affleck, M. Dine and N. Seiberg,
``Dynamical Supersymmetry Breaking in Four Dimensions and its
Phenomenological Implications'', \np{256}{1985}{557}.}%
\lref\ismir{K. Intriligator and N. Seiberg, ``Mirror Symmetry in Three
Dimensional Gauge Theories,'' hep-th/9607207, \pl{387}{1996}{513}.}
\lref\gl{J. P. Gauntlett and D. A. Lowe, ``Dyons and S-Duality in
$N=4$ Supersymmetric Gauge Theory,'' hep-th/9601085,
\np{472}{1996}{194}.} 
\lref\lwy{K. Lee, E.J. Weinberg, and P. Yi, ``The Moduli Space of Many
BPS Monopoles for Arbitrary Gauge Groups,'' hep-th/9602167,
\physrev{54}{1996}{1633}.}
\lref\swtd{N. Seiberg and E. Witten, ``Gauge Dynamics and
Compactification to Three Dimensions,'' hep-th/9607163.}
\lref\kv{S. Katz and C. Vafa, ``Geometric Engineering of $N=1$ Quantum
Field Theories,'' hep-th/9611090.}
\lref\ew{E.J. Weinberg, ``Fundamental Monopoles and Multimonopole
Solutions for Arbitrary Simple Gauge Groups'', \np{167}{1980}{500}.}
\lref\seibergfive{N. Seiberg, ``Five Dimensional SUSY Field Theories,
Non-Trivial Fixed Points and String Dynamics'', hep-th/9608111.}%
\lref\polyakov{A. M. Polyakov, ``Quark Confinement and the Topology of
Gauge Groups'', \np{120}{1977}{429}.}%
\lref\waddas{S.R. Wadia and S.R. Das, ``Topology of Quantum Gauge
Fields and Duality. 1. Yang-Mills Higgs System in (2+1) Dimensions'',
\pl{106}{1981}{386}.}
\lref\jr{R. Jackiw and C. Rebbi, ``Solitons with Fermion Number
$1/2$'', \physrev{13}{1976}{3398}.}%
\lref\callias{C. Callias, ``Axial Anomalies and 
Index Theorems on Open Spaces'', \cmp{62}{1978}{213}.}%
\lref\leeyi{K. Lee and P. Yi, ``Monopoles and Instantons in Partially
Compactified D-branes'', hep-th/9702107.}%
\lref\gomez{C. G\'omez and R. Hern\'andez, ``M and F-Theory
Instantons, $N=1$ Supersymmetry and Fractional Topological Charge'',
hep-th/9701150.}%
\lref\hw{A. Hanany and E. Witten, ``Type IIB Superstrings, BPS
Monopoles, and Three Dimensional Gauge Dynamics'', hep-th/9611230.}%
\lref\bdl{M. Berkooz, M. R. Douglas and R. G. Leigh, ``Branes
Intersecting at Angles'', hep-th/9606139, \np{480}{1996}{265}.}%
\lref\egk{S. Elitzur, A. Giveon and D. Kutasov, ``Branes and N=1
Duality in String Theory'', hep-th/9702014.}%
\lref\bhoy{J. de Boer, K. Hori, Y. Oz and Z. Yin, ``Branes and Mirror
Symmetry in $N=2$ Supersymmetric Gauge Theories in Three Dimensions'',
hep-th/9702154.}%
\lref\redlich{A.N. Redlich, ``Gauge Noninvariance and Parity Violation of
Three-Dimensional Fermions,'' \prl{52}{1984}{18}\semi
A.N. Redlich,
``Parity Violation and Gauge Non-invariance of the Effective Gauge Field
Action in Three Dimensions,'' \physrev{29}{1984}{2366}.}
\lref\wittanom{E. Witten, ``An $SU(2)$ Anomaly,'' \pl{117}{1982}{324}.}
\lref\agw{L. Alvarez-Gaum\'e and E. Witten, ``Gravitational Anomalies,''
\np{234}{1983}{269}.}
\lref\swfour{N. Seiberg and E. Witten, 
``Electric-Magnetic Duality, Monopole
Condensation and Confinement in $N=2$ Supersymmetric Yang-Mills
Theory'', \np{426}{1994}{19}, hep-th/9407087\semi
N. Seiberg and E. Witten, ``Monopoles, Duality and Chiral
Symmetry Breaking in $N=2$ Supersymmetric QCD'', \np{431}{1994}{484},
hep-th/9408099.}%
\lref\acharya{B.S. Acharya, ``A Mirror Pair of Calabi-Yau Fourfolds in
Type II String Theory'', hep-th/9703029.}%
\lref\bhoo{J. de Boer, K. Hori, H. Ooguri and Y. Oz, ``Mirror Symmetry
in Three Dimensional Gauge Theories, Quivers and D-branes'',
hep-th/9611063.}%
\lref\bhooy{J. de Boer, K. Hori, H. Ooguri, Y. Oz and Z. Yin, ``Mirror
Symmetry in Three Dimensional Gauge Theories, $SL(2,\IZ)$ and D-brane
Moduli Spaces'', hep-th/9612131.}%
\lref\porrati{M. Porrati and A. Zaffaroni, ``M Theory Origin of Mirror
Symmetry in Three Dimensional Gauge Theories'', hep-th/9611201.}%
\lref\nonren{N. Seiberg, ``Naturalness Vs. Supersymmetric
Non-renormalization Theorems'', hep-ph/9309335, \pl{318}{1993}{469}.}%
\lref\bpsbound{Z. Hlousek and D. Spector, ``Why Topological Charges
Imply Extended Supersymmetry'', \np{370}{1992}{143}\semi
J. Edelstein, C. N\'u\~nez and F. Schaposnik, ``Supersymmetry and
Bogomolny Equations in the Abelian Higgs Model'', hep-th/9311055,
\pl{329}{1994}{39}.}%
\lref\bogo{E.B. Bogomolny, ``The Stability of Classical Solutions,''
Sov. J. Nucl. Phys. {\bf 24} (1976) 449.}
\lref\nielol{
H.B. Nielsen and P. Olesen, ``Vortex -- Line Models for Dual
Strings,'' \np{61}{1973}{45}.}
\lref\twodinst{ P. Di Vecchia and S. Ferrara, ``Classical Solutions in
Two-Dimensional Supersymmetric Field Theories,'' \np{130}{1977}{93}.}
\lref\fourdstrng{A. Hanany
and I.R. Klebanov, ``On Tensionless Strings in (3+1) Dimensions'',
hep-th/9606136, \np{482}{105}{1996}; B.R. Greene, D.R. Morrison and
C. Vafa, ``A Geometric Realization of Confinement'', hep-th/9608039,
\np{481}{1996}{513}.}
\lref\barbon{J.L.F. Barbon, ``Rotated Branes and $N=1$ Duality'',
hep-th/9703051.}%
\lref\intseirev{K. Intriligator and N. Seiberg, ``Lectures on
Supersymmetric Gauge Theories and Electric-Magnetic Duality'',
hep-th/9509066, {Nucl. Phys. Proc. Suppl. {\bf 45BC} (1996) 1}.}%
\lref\peskin{M. E. Peskin, ``Duality in Supersymmetric Yang-Mills
Theory'', hep-th/9702094.}%
\lref\power{N. Seiberg, ``The Power of Holomorphy : Exact Results in
4D SUSY Field Theories'', hep-th/9408013, PASCOS 1994:357-369.}%
\lref\irdyn{N. Seiberg, ``IR Dynamics on Branes and Space-Time
Geometry'', hep-th/9606017, \pl{384}{1996}{81}.}%
\lref\gates{H. Nishino and S. J. Gates, Jr., ``Chern-Simons Theories
with Supersymmetries in Three Dimensions'', \ijmp{8}{1993}{3371}.}%
\lref\berkman{J. de Boer, K. Hori, H. Ooguri, Y. Oz, and Z. Yin, 
``Mirror Symmetry in Three Dimensional Gauge Theories, $SL(2,\IZ)$, 
and $D$-Brane Moduli Spaces,'' hep-th/9612131.}
\lref\berknew{J. de Boer, K. Hori, and Y. Oz, 
``Dynamics of $N=2$ Supersymmetric Gauge Theories in Three
Dimensions,'' hep-th/9703100.}
\def\FI{Fayet-Iliopoulos}

\Title{hep-th/9703110, RU-97-10, IASSNS-HEP-97/18}
{\vbox{\centerline{Aspects of $N=2$ Supersymmetric Gauge}
       \centerline{Theories in Three Dimensions}}}
\medskip
\centerline{O. Aharony$^1$, A. Hanany$^2$, 
K. Intriligator\footnote{${}^*$}{On leave 1996-1997
{}from Department of Physics, University of California, San
Diego.}$^2$, N. Seiberg$^1$, and M.J. Strassler$^{2,3}$}
\vglue .5cm
\centerline{$^1$Dept. of Physics and Astronomy,}
\centerline{Rutgers University}
\centerline{Piscataway, NJ 08855, USA}
\vglue .3cm
\centerline{$^2$School of Natural Sciences}
\centerline{Institute for Advanced Study}
\centerline{Princeton, NJ 08540, USA}
\vglue .3cm
\centerline{$^3$Department of Physics}
\centerline{Harvard University}
\centerline{Cambridge, MA 02138, USA}
\medskip
\noindent

We consider general aspects of $N=2$ gauge theories in three
dimensions, including their multiplet structure, anomalies and
non-renormalization theorems. For $U(1)$ gauge theories, we discuss
the quantum corrections to the moduli space, and their relation to
``mirror symmetries'' of 3d $N=4$ theories.  Mirror symmetry is given
an interpretation in terms of vortices.   For $SU(N_c)$ gauge groups with
$N_f$ fundamental flavors, we show that, depending on the number of
flavors, there are quantum moduli spaces of vacua with various
phenomena near the origin.

\Date{3/97}                                   

%\draftmode          

\newsec{Introduction}

Recent exact results in four dimensional supersymmetric theories
have given insight into the
quantum dynamics of strongly coupled gauge theories, revealing
interesting phenomena and phases (for reviews see, e.g.\
\refs{\intseirev,\peskin}). Interesting phenomena have also been
found recently for three dimensional $N=4$ theories, in particular
mirror symmetry which relates the IR behavior of two different field
theories, interchanging their Higgs and Coulomb branches \ismir.
Additional examples of mirror symmetry 
and connections to string theory appeared in 
\refs{\bhoo,\porrati,\hw, \berkman}.
It is interesting to see how these
phenomena depend on the dimension and the amount of supersymmetry, and
how they behave upon compactification.  Another
motivation is that there are powerful connections between gauge
theories in various dimensions and the dynamics of string theory.  New
information about field theory can lead to new insight into string
theory and vice-versa. The analysis of three dimensional field
theories may also have applications for statistical mechanics
problems, which are beyond the scope of this paper.

We consider gauge theories in $d=3$ dimensions with four
supercharges, the same number as for $N=1$ supersymmetry in $d=4$,
which corresponds to $N=2$ supersymmetry in $d=3$. In string theory,
such theories arise as the low-energy theories of compactifications of
M theory on Calabi-Yau fourfolds\foot{Mirror symmetry in this context
and its relation to mirror symmetry in string theory were 
recently discussed in \acharya.} (as well as {}from dimensional
reduction of $d=4$ $N=1$ theories). They also correspond to the
low-energy field
theories of membranes in M theory compactifications on fourfolds or on
threefolds, and of D2-branes in type IIA compactifications on
threefolds. We will not discuss these relations to string theory here.

Theories with $N=1$ supersymmetry in three dimensions have no
holomorphy properties, so we cannot control their non-perturbative
dynamics. However, as we will discuss, with $N=2$
supersymmetry there are holomorphic objects and non-renormalization
theorems, which enable us to compute some properties of these theories
exactly.

To summarize our results, we find exact superpotentials in $U(1)$,
$SU(N_c)$, and $U(N_c)$ examples.  Perturbative effects in Abelian
theories can cause the Coulomb branch to split into several regions,
described by different variables.  At the intersections of different
regions are RG fixed points for which we find dual descriptions.  In
non-Abelian cases, instantons (analogous to $d=4$ monopoles) can lift
most or all of the Coulomb branch, and sometimes also the Higgs
branch. In 3d $N=2$ $SU(N_c)$ SQCD with $N_f<N_c-1$ flavors, for
example, the moduli space is completely lifted and the theory has no
stable vacuum.  In some cases there is a quantum moduli space with the
classically distinct Higgs and Coulomb branches smoothly merged
together.  This happens, for example, in $SU(N_c)$ SQCD theories with
$N_f=N_c-1$ flavors. In other cases, both for Abelian and Non-Abelian
theories, there are distinct Higgs and Coulomb branches, with RG fixed
points where they intersect.  In some cases we find dual descriptions
of these fixed points.

We begin in sect. 2 by discussing some general classical and quantum
aspects of 3d $N=2$ theories.  For example, we discuss linear
multiplets, central charges, \FI\ terms, and non-renormalization
theorems.  We point out that the ``parity anomaly'' of 
\refs{\redlich, \agw} can be used to give an analog of the 't Hooft
anomaly matching condition for 3d theories.  

In sect. 3 we analyze the dynamics of $U(1)$ gauge theories.  For
$N_f>0$, the moduli space consists of different branches which meet at
a point at the origin, where there is a RG fixed point.  For $N_f=1$,
we find a dual description of the RG fixed point and verify that our
discrete anomaly matching conditions are satisfied.

In sect. 4 we discuss dual descriptions of the RG fixed points at the
origin of $N=2$ SQED with $N_f>0$ which are obtained {}from the ``mirror
symmetry'' duality of  the corresponding $N=4$ theories \ismir.  We
connect these duals with the results of sect. 3.

In sect. 5 we discuss vortices in $N=4$ and $N=2$ theories and 
their connection with mirror symmetry of $N=4$ and $N=2$ SQED.

In sect. 6 we analyze the dynamics of the $SU(2)$ gauge theory with
various numbers of quark flavors and mass terms.  For $N_f=1$ there is
a quantum moduli space of vacua, with the classically distinct Higgs
and Coulomb branches smoothly merged together.  For $N_f\geq 2$ there
are distinct Higgs and Coulomb branches which intersect at a point at
the origin, where there is a RG fixed point.  For $N_f=2$ we find a
dual description of the RG fixed point at the origin.  We discuss
adding real mass terms, connecting with our results for $U(1)$
theories.

In sect. 7 we analyze $d=4$ $N=1$ $SU(2)$ gauge theories compactified
on a circle of varying radius, showing how known non-perturbative
effects in the 4d theories can be related to the quantum effects we
found in 3d theories.

In sect. 8, we generalize the discussion of sect. 6 and sect. 7
to $SU(N_c)$ gauge theories. 

{\bf Note added:} As we completed this paper, we received \berknew,
which considers the same theories.

\newsec{General aspects of $d=3$ $N=2$ theories}

$N=2$ supersymmetry in three dimensions has four supercharges, with an
algebra which follows simply {}from reducing $d=4$ $N=1$ supersymmetry
down to three dimensions: 
\eqn\algebra{\{Q_\alpha , Q_\beta\} =\{ \overline
Q_\alpha,
\overline Q_\beta \}=0,\qquad \{Q_\alpha , \overline Q_\beta \}=2\sigma
^{\mu}_{\alpha \beta }P_{\mu}+2i\epsilon _{\alpha \beta}Z,} where the
$\sigma^{\mu}$ are chosen to be real and symmetric.  $Z$ is a real
central term, which in the dimensional reduction corresponds to the
momentum $P_3$ in the reduced direction. The spinors $Q$ and
$\overline Q$ are complex, and thus include twice the minimal amount
of charges in three dimensions.  As in four dimensions, the
automorphism of this algebra is $U(1)_R$, rotating the supercharges.

As in four dimensions, there are chiral superfields $X$ which satisfy
$\overline D_\alpha X=0$, anti-chiral superfields $\overline X$
which satisfy $D_\alpha {\overline X}=0$, and vector superfields which
satisfy $V = V^{\dagger}$. In addition, in three dimensions there can be
linear multiplets $\Sigma$, which satisfy $\epsilon ^{\alpha
\beta}D_\alpha D_\beta \Sigma=\epsilon ^{\alpha \beta}\overline
D_\alpha \overline D_\beta \Sigma =0$, whose lowest component is a
real scalar field. Occasionally, we will find it convenient to dualize
the vector and linear multiplets into chiral multiplets.

All states satisfy a BPS bound of the form $M\geq |Z|$
\bpsbound. General irreducible representations of the above algebra,
such as the chiral and vector representations, contain two (real)
bosonic and two (Majorana) fermionic degrees of freedom. Smaller
representations with one real bosonic degree of freedom and one
fermion can also exist, and must saturate the BPS bound $M =
|Z|$. However, as discussed in \S2.3, non-zero $Z$ can only occur for
states which are charged under $U(1)$ symmetries, and then CPT
dictates the existence also of a conjugate representation.

\subsec{Wess-Zumino theories in $d=3$}

Consider a Wess-Zumino type Lagrangian for chiral and anti-chiral
superfields $X$ and $\overline X$. The general Lagrangian we get by
reduction {}from $d=4$ is of the form
\eqn\genwz{\int d^4\theta ~K(X, \overline X)
+(\int d^2\theta ~W(X)+h.c.).}

A three dimensional chiral superfield $X$ has engineering dimension
$1/2$, and thus the classically marginal superpotential is $W=X^4$,
corresponding to an $|X|^6$ scalar potential.  Unlike four dimensions,
where WZ theories always flow to Gaussian fixed points, in three
dimensions the WZ theory with superpotential $W=X^3$ flows to an
interacting fixed point in the infrared.  It follows {}from the three
dimensional superconformal algebra that, as in
\power, for any $N=2$ theory the
dimensions of all operators satisfy
\eqn\dineq{D\geq |R|,}
where $R$ is the charge under the $U(1)_R$ symmetry which is in the same
multiplet as the stress tensor.  The inequality \dineq\ is saturated
for chiral and anti-chiral operators.  For $W=X^3$, $R(X)=2/3$ and
thus \dineq\ gives 
\eqn\exactad{D(X)={2\over 3},} 
giving the {\it exact} anomalous dimension of $X$ at the infrared
fixed point.

\subsec{Gauge theories}

The vector multiplet $V$ of $d=3$ $N=2$ supersymmetry contains, in
addition to the $d=3$ vector potential, a real scalar $\phi$ in the
adjoint of the gauge group $G$; $\phi$ corresponds to the component of
the $d=4$ vector potential in the reduced direction.  The massless
vector multiplet also contains two real (which can be joined into one
complex) fermion gauginos.  As in four dimensions, there are chiral
and anti-chiral field strengths, defined as $W_\alpha = -{1\over 4}
{\overline D}{\overline D} e^{-V} D_{\alpha} e^V$ and ${\overline
W}_{\alpha} = -{1\over 4} D D e^{-V} {\overline D}_{\alpha} e^V$, and
the classical gauge kinetic terms, including a kinetic term for the
real scalar $\phi$, are
\eqn\clgkt{{1\over g^2}\int d^2\theta ~\Tr ~W_{\alpha}^2 + h.c.}
The gauge field and real scalar are neutral under the $U(1)_R$
symmetry, while the complex fermion has a charge we will normalize to
be $+1$.

When the gauge group $G$ is Abelian or has Abelian factors, it is
possible also to add a \FI\ term.  As in 4d, the \FI\ term is of the form 
$\zeta \int d^4\theta V$ for some real \FI\ parameter $\zeta$.

The moduli space of vacua has a ``Coulomb branch'' where the real
scalar $\phi$ gets an expectation value in the Cartan subalgebra of the
gauge group, breaking $G$ to the Cartan subgroup $U(1)^r$, with
$r=$rank$(G)$.  The Coulomb branch is thus a Weyl chamber, which is a
{\it wedge}\/ subspace of $\IR ^r$ parameterized by Cartan scalars
$\phi ^j$ in $\IR ^r/\CW$ ($\CW$ is the Weyl group of $G$), the
expectation values of the scalars in the
massless Cartan $U(1)^r$ vector multiplets
$V ^j$.  At the boundaries of the Weyl chamber, at the classical
level, there is enhanced gauge symmetry.

In the bulk of the Coulomb branch, the $U(1)^r$ gauge fields can be
dualized to scalars via $F ^{(j)}_{\mu
\nu}=\epsilon _{\mu\nu\sigma}\partial ^\sigma \gamma ^j$, $j=1\dots r$.  
Due to charge quantization, the scalars $\gamma^j$ live on an
$r$-dimensional torus. It is the Cartan torus of the dual gauge group,
whose size is of the order of the gauge coupling $g$ \ref\nsta{
N. Seiberg, ``Notes on theories with 16 supercharges,'' RU-97-7, to
appear.}.  The currents $J_{\mu}^{(j)} = \epsilon_{\mu \nu
\rho} (F^{\nu \rho})^{(j)}$ generate ``magnetic'' $(U(1)_J)^r$ 
global symmetries, corresponding to shifts of $\gamma^j$; these
symmetries are exact only for Abelian gauge groups.  The $\gamma ^j$
can be combined with the $\phi ^j$ into chiral superfields $\Phi ^j$,
with scalar component $\Phi ^j=\phi ^j+i\gamma ^j$.  At the boundaries
of the Coulomb wedge, where the gauge group is classically
non-Abelian, there is no known way to dualize the gauge fields into
scalars.  In the quantum theory, these are the regions where we expect
quantum effects to be especially important.

There can also be matter multiplets $Q_f$ in representations $R_f$ of
the gauge group.  The classical Lagrangian has terms 
\eqn\matcl{\sum _f\int d^4\theta ~Q_f^\dagger e^V Q_f,}
where $V$ includes a term $\phi \theta \overline{\theta}$ for the real
adjoint scalar $\phi$.  In particular, the Lagrangian includes a
potential for the squarks of the form
\eqn\mvterm{\sum _f |\phi Q_f|^2,}
and $\ev{\phi}$ looks like a ``real mass'' for the matter fields (we
will discuss this further in \S2.3).

In addition to the Coulomb branch, there can be a ``Higgs'' branch,
where the squark components of the matter multiplets $Q_f$ get
expectation values.  Because of the coupling \mvterm, $\ev{Q_f}\neq 0$
generally requires $\ev{\phi}=0$ (at least for some of the components
of $\phi$), and vice-versa.  Therefore, the complete classical moduli
space of vacua generally consists of distinct branches, Coulomb with
$\ev{\phi}
\neq 0$ and $\ev{Q_f}=0$, and Higgs with $\ev{Q_f}\neq 0 $ and $\ev{\phi
}=0$.  This is similar to $N=2$ supersymmetric theories in four
dimensions.  As is the case there, depending on the matter content
there can also be mixed branches where some $\ev{Q_f}$ and $\ev{\phi}\neq
0$ such that \mvterm\ still vanishes.  

There is a freedom in choosing the $U(1)_R$ charge assignments of the
matter multiplets, as $U(1)_R$ can mix with other $U(1)$ global
symmetries which act on the matter supermultiplets. Thus, generally it
will not be simple to determine the dimensions of chiral fields at IR
fixed points via \dineq, since we do not know which $U(1)_R$ current
is in the same multiplet as the stress tensor there. In three
dimensions there is no condition on the symmetries that they be
anomaly free.  For convenience, we will choose the squarks to be
neutral under $U(1)_R$; the quarks then have charge $(-1)$.

As discussed in \ahw, non-perturbative effects can lead to a
dynamically generated superpotential which lifts the classical moduli
space degeneracy of the Coulomb branch.  In addition, when there is
also a Higgs branch, the intersection of the branches near the origin
of the moduli space of vacua, where quantum effects are strong, can be
interesting.  We will discuss such effects in this paper.

\subsec{Linear multiplets, central charges, Fayet-Iliopoulos terms and
non-renormalization theorems}

The vector multiplet in $d=3$ $N=2$ theories includes a vector, a
scalar and two Majorana fermions. In \S2.2, we defined the chiral
superfield $W_\alpha$ whose lowest component is a gaugino, and wrote
the gauge kinetic term in terms of this superfield, which is
gauge-invariant for Abelian theories (we will only discuss Abelian
vector fields in this sub-section). However, we can also define a
superfield whose lowest component is the scalar in the vector
multiplet -- this is a linear multiplet\foot{These multiplets were
called ``scalar field strengths'' in \gates.}, defined by $\Sigma =
\epsilon^{\alpha \beta} {\overline D}_{\alpha} D_{\beta} V$. 
It is easy to check that this $\Sigma$ satisfies 
$D^2 \Sigma = {\overline D}^2 \Sigma = 0$ (where
$D^2 = \epsilon^{\alpha \beta} D_{\alpha} D_{\beta}$), and that it is
gauge invariant (under $V \to V + i (\Lambda - \Lambda^{\dagger})$). 
The lowest component of $\Sigma$ is the scalar
$\phi$, and its expansion includes also a term ${\overline \theta}
\sigma_\rho \theta F_{\mu \nu} \epsilon^{\rho \mu \nu}$. 
The gauge kinetic term may be written simply in the
form $\int d^4\theta \Sigma^2$. In three dimensions, we can dualize
the vector into a scalar, and turn the linear (or vector)
multiplets into chiral multiplets. Note, however, that chiral
multiplets obtained in this way always have $U(1)_J$ symmetries
which act as shifts by an imaginary number on their scalar component.

Linear multiplets are useful also in describing global conserved
currents. The conservation of the global currents, which follows {}from
the equations of motion, implies that they can be viewed as components
of linear multiplets, satisfying $D^2 J = {\overline D}^2 J = 0$. In
fact, the multiplet $\Sigma$ we defined above is an example of this
phenomenon, and it includes the conserved current $J^{\mu} =
\epsilon^{\mu \nu \rho} F_{\nu
\rho}$, which generates the global $U(1)_J$ symmetry. Generally, for a
conserved current $J_\mu$, there will be a linear multiplet $J$ which
includes the term ${\overline \theta} \sigma^{\mu} \theta J_{\mu}$.
However, since the current is only conserved on-shell,
we cannot always find a vector superfield $V$ such that $J =
\epsilon^{\alpha \beta} {\overline D}_{\alpha} D_{\beta} V$.

The SUSY algebra includes a real central charge $Z$, appearing in 
\algebra.  In 4d $N=2$ theories, $Z$ can get contributions {}from 
both local and global currents \swfour.  However, in 3d the mass (and
the central charge) of particles charged under local currents are
divergent, even classically, due to the $1/r$ fall-off of electric
fields in 3d.  Thus, we cannot really discuss the masses of
electrically charged fields, and we can only use the BPS formula for
gauge-neutral states.  Therefore, only global charges contribute to
the central charge of physical states.

A simple example of a theory with a non-zero central charge is given
by the Lagrangian 
\eqn\realmass{\int d^4 \theta X^{\dagger} e^{\tilde m \theta
\overline{\theta}} X,}
where $X$ is a chiral multiplet.  When written in components this
Lagrangian includes terms of the form $(\tilde m^2 |X|^2+i\tilde m
\epsilon ^{\alpha
\beta}\overline{\psi}_\alpha \psi_\beta )$. The parameter $\tilde m$
appears here as a ``real mass'' term for $X$, which is distinct {}from
the complex masses appearing in the superpotential. The central charge
$Z$ (if we promote it to a background superfield)
corresponds to the linear multiplet $J$ containing the global
current under which $X$ is charged. Its scalar component
is given by $Z = \tilde m$, and $X$ saturates the BPS bound. 
As discussed above, chiral
multiplets do not have to saturate the BPS bound. For instance, if we
add to \realmass\
a superpotential of the form $W = m X Y$ (where $Y$ is another
chiral superfield needed to preserve the global $U(1)$),
the (tree-level) mass of $X$ is $M =
\sqrt{\tilde{m}^2 + |m|^2}$, and the BPS bound is no longer saturated.

More generally, $Z$ gets contributions {}from global Abelian currents,
of the form \eqn\sqm{Z = \sum q_i m_i,} where $q_i$ is the charge of
the field (or the state) under a global $U(1)_i$ symmetry, and $m_i$
is a parameter. As in \nonren, we promote every parameter to a
background superfield. The parameters $m_i$ in \sqm\ are then in
background vector or linear multiplets. To see this, add to any system
with a global $U(1)$ symmetry a charged chiral multiplet as in
\realmass\ with a large $U(1)$ charge. The large charge makes this
field heavy and almost decoupled.  Its mass is given by the BPS
formula as its charge times $m$ where $m$ is the scalar of a linear
multiplet. Hence, even without this heavy auxiliary state, the
coefficient in $Z$ is in a linear multiplet.  This sort of
``background gauging'' can only be performed if the global symmetry is
exact, and only then will this type of ``real mass'' terms appear.
The fact that we cannot write them down in general is consistent with
the fact that we have no simple superspace expression for these terms.
(This is also related to the fact that $J$ is only in a linear
multiplet when we use the equations of motion, and that generally we
cannot view it as originating {}from a vector multiplet.)

Another example of a global symmetry which can contribute to the
central charge is the $U(1)_J$ symmetry, which corresponds to shifting
the dual photon $\gamma$. This is only a symmetry in Abelian theories.
The current multiplet $J$ is exactly the linear multiplet
$\Sigma$ defined above, and the added term is of the form $\int d^4
\theta V_b \Sigma = \int d^4
\theta \Sigma_b V$ by integrating by parts, where $\Sigma_b =
\epsilon^{\alpha \beta} {\overline D}_{\alpha} D_{\beta} V_b$
is a background linear multiplet. Thus, the scalar component of the
background vector field is exactly the
\FI\ term $\zeta$, and $Z$ will get a contribution in \sqm\ with 
$m_J=\zeta$.   States with charge $q_J$ under $U(1)_J$
thus obey a BPS bound of the form $M \geq |q_J\zeta|$.

In the spirit of \nonren, the results of this section suggest two
types of (non-perturbative) non-renormalization theorems for these
theories. First, it is easy to see that (to preserve supersymmetry)
the superpotential cannot include linear multiplets, but only chiral
multiplets, and, therefore, it cannot depend on the ``real mass''
terms or on \FI\ terms (which are all background linear
multiplets). Alternatively, if we dualize the linear multiplets into
chiral multiplets, holomorphy together with the shift symmetry
mentioned above forbid these multiplets {}from appearing in the
superpotential (as long as the shift symmetry is exact).  Second, we
found that the central charge $Z$ is a background linear multiplet.
Any dependence on chiral multiplets, or any non-linear dependence on
linear multiplets, would change this fact. Thus, the relation \sqm\
between $Z$ and the ``real mass'' terms and \FI\ parameter is, in
fact, exact.  This is analogous to the similar phenomenon in $d=4$
$N=2$ gauge theories \swfour, where the central charges are exactly
equal to the scalar components of $N=2$ vector multiplets.

\subsec{Anomalies}

Unlike the situation in four dimensions, 
in three dimensions there are no local gauge anomalies.
However, as found in \refs{\redlich, \agw}, gauge invariance can
require the introduction of a classical Chern-Simons term, which
breaks parity.  This is referred to as a ``parity anomaly.''

Consider first an Abelian $U(1)^r$ gauge theory.  There can be
classical Chern-Simons couplings,
\eqn\abcs{\sum _{i,j=1}^rk_{ij}\int d^4\theta ~\Sigma _i V_j,}
where $\Sigma _i=\epsilon ^{\alpha \beta} {\overline D}_\alpha D_\beta
V_i$ are linear superfields; \abcs\ is the supersymmetric completion
of $\sum _{ij} k_{ij} A_i\wedge dA_j$.  
We work in a basis for $U(1)^r$ where all charges are integers.
At the
quantum level, if we integrate out the charged fermions,
there is an additional induced contribution to the
Chern-Simons term, coming {}from a one-loop diagram with charged
fermions running in the loop:  
\eqn\kcsiout{(k_{ij})_{eff}=k_{ij}+\half \sum _f (q_f)_i(q_f)_j {\rm
sign} (M_f);} the sum runs over all fermions, $(q_f)_i$ is the charge
of fermion $f$ under $U(1)_i$ (in units of the quantized charge), and
$M_f$ is the mass of fermion $f$.  In the present context (without a
superpotential), $M_f=\tilde
m_f+ \sum_{i=1}^r (q_f)_i\phi _i$.

Gauge invariance restricts the coefficients of the Chern-Simons term
as $(k_{ij})_{eff}\in \IZ$.  {}From \kcsiout, we see that the bare
Chern-Simons coefficients $k_{ij}$ must satisfy the quantization
conditions
\eqn\kabquant{k_{ij}+\half \sum _f(q_f)_i(q_f)_j~\in \IZ,}
where $f$ runs over all fermions.  In particular, when $\sum _f
(q_f)_i(q_f)_j$ is odd, $k_{ij}\neq 0$ and parity is necessarily
broken.  This is the ``parity anomaly.''  Without the classical
Chern-Simons term, the fermion determinant in these cases would be
multiplied by $(-1)$ under certain gauge transformations.  The $k_{ij}
\in \IZ +\half$ Chern-Simons term 
plays the role of a Wess-Zumino term whose lack of gauge invariance
compensates for that of the fermion determinant.

If we introduce background gauge fields for the global $U(1)$
symmetries, as described in \S2.3, and integrate out the fermions,
similar Chern-Simons terms will appear (at one-loop)
involving these background
gauge fields as well. In particular, terms of the form $\int
d^4\theta \Sigma_b V$ where $\Sigma_b$ is a background linear
multiplet and $V$ is a gauge field can be generated (by the same
diagrams which generate \kcsiout), which correspond
to \FI\ terms for the gauge field $V$. When such terms appear, the
global symmetries corresponding to $\Sigma_b$ will be mixed with the
$U(1)_J$ symmetry corresponding to $V$, and the chiral superfields
corresponding to the dual photons will transform non-trivially under
the other global symmetries as well. Below we will see this phenomenon
in several examples.

There is a similar parity anomaly for non-Abelian theories
\refs{\redlich, \agw}.  Just as the
4d anomaly of \wittanom\ is associated with $\pi _4(G)$, the 3d parity
anomaly is associated with $\pi _3(G)$.  Under the large gauge
transformations corresponding to non-trivial elements of $\pi _3(G)$,
the fermion determinant can pick up a minus sign, making the theory
inconsistent.  Unlike the situation in 4d, however, in 3d this anomaly
can always be cancelled by adding a Chern-Simons term with a
coefficient $k$ which is half-integral \refs{\redlich, \agw}.  Again,
the Chern-Simons term with half-integral $k$ plays the role of a
Wess-Zumino term whose lack of gauge invariance compensates for that
of the fermion determinant.  The condition on the classical
Chern-Simons term $k$ is
\eqn\nonabk{k+\half \sum _fd_3(R_f)\in \IZ,}
where the sum is over all fermions $f$ in representations $R_f$ of $G$
and $d_3(R_f)$ is the cubic index of $R_f$, normalized so that the
${\bf N}$ of $SU(N)$ has $d_3({\bf N})=1$.
In particular, when $\sum _f d_3(R_f)$ is odd, $k\neq 0$ and parity is
necessarily broken; there is a ``parity anomaly.''

It is interesting to note that the parity anomaly gives an analog of
the 't Hooft anomaly matching conditions in 3d.  In 4d, the precise
anomalies associated with gauging global symmetries must match between
the microscopic and the low energy theories. In 3d, there is a weaker
$\IZ _2$ type condition: whether or not the gauged global symmetry
would have a parity anomaly must match between the microscopic theory
and any other theory which is equivalent to it in the IR.

The Chern-Simons term gives the photon a mass, lifting the Coulomb
branch discussed earlier. In most of this paper we will not be
interested in that situation, and will want to avoid turning on the
Chern-Simons term.  The matter content must then be chosen so that
there is not a parity anomaly.

\subsec{Instantons}
 
In three dimensions, 
instantons are associated with $\pi _2$.  Because $\pi _2(G)=0$
for any gauge group, there can only be instantons on the Coulomb
branch of non-Abelian theories, 
where the gauge group is broken to the Cartan torus $U(1)^r$.
In that case, the relevant configurations are related to elements of
$\pi _2(G/U(1)^r)= \IZ
^r$.  (Also, because $\pi _2(G)=0$, there is no analog of the 4d theta
angle in three dimensions.)  It follows {}from $\pi _2(G/U(1)^r)= \IZ
^r$ that there are $r=$rank$(G)$ independent fundamental instantons in
$d=3$, associated with the simple roots of the gauge group.  This is
the same as the $d=4$ result that there are $r$ independent monopoles
\ew\ (which is not surprising since the $d=3$ instantons are identical
to $d=4$ monopoles when ignoring the $d=4$ time dimension). The $j$'th
``fundamental'' instanton is semi-classically weighted by
$e^{-S_j}=e^{-\phi \cdot \beta _j/g^2}$, where $\phi$ is the adjoint
scalar (in the Cartan subalgebra) defined above, and $\beta_j$,
$j=1,\dots, r$, are the correct basis of simple roots discussed in
\ew.  In a given Weyl chamber, the $\beta _j$ will be 
chosen so that $\phi \cdot \beta_j
\geq 0$ for all $j$;   at the boundary of the Weyl
chamber where a given $\phi \cdot \beta _j=0$, there is an enhanced
$SU(2)_j \subset G$. Instantons corresponding to linear
combinations of the $\beta _j$ have additional zero modes
corresponding, as with monopoles in $d=4$ \ew, to separating them into
linear combinations of the $r$ fundamental instantons. 

It is natural to combine the above instanton factors 
with the dualized photons, which are similar to theta angles
in $d=4$, to make quantities holomorphic in $\Phi ^k=\phi ^k+i\gamma ^k$,
\eqn\monj{Y_j\sim e^{\Phi \cdot \beta _j/g^2}.}
The fields $Y_j$ are normalized so that they have charge $+1$ under
the (approximate) $U(1)_J$ symmetries that these theories have on
their Coulomb branch. They provide a natural set of coordinates for
the Coulomb branch; the sign of the exponent in \monj\ is chosen so
that large $Y_j$ corresponds to being far out along the Coulomb
branch.  The $j$-th instanton contribution is weighted by $Y_j^{-1}$,
including correctly the dependence on the dual photon \polyakov.  The
$\sim$ in \monj\ is because this relation between $Y_j$ and $\Phi$ is
valid only semi-classically, for large $\phi\cdot\beta_j$; it can be
modified near the boundaries of the Coulomb branch.

As discussed in \S2.4, the fields $Y_j$ can acquire charge under
global symmetries due to one-loop effects. This is also consistent
with the counting of fermionic zero modes in the instanton
background. Then, $\ev{Y_j}$ spontaneously breaks the global symmetry,
with the dualized photon playing the role of the Goldstone boson \ahw.
The fermion zero modes associated with the instantons are easily
obtained: the quark zero modes are the same as the $d=4$ fermionic
zero modes in a monopole background
\refs{\jr,\callias}, while the gluino zero modes 
are related by supersymmetry to the bosonic zero modes which were
analyzed in \ew.  It thus follows that the $j$-th ``fundamental''
instanton always has
two gaugino zero modes (this is half of the number
of zero modes for the $N=4$ theories analyzed in \swtd, since we have
half the number of gluinos).  
In particular, in pure $d=3$ $N=2$ Super-Yang-Mills 
theory with no matter fields,
each fundamental instanton has two gaugino zero modes (and no other
fermionic zero modes), so the $Y_j$ all
carry charge $(-2)$ under the global $U(1)_R$ (gaugino number) symmetry.

\newsec{$U(1)$ gauge theories}

Consider $U(1)$ gauge theory with $N_f$ ``flavors'',
$Q^i,\tilde{Q}_{\tilde i}$ ($i, \tilde i=1,\cdots,N_f$), with charge
$\pm 1$.  Unlike the situation in 4d, where this theory is IR free, in
3d it has interesting dynamics.  Before taking into account the
quantum corrections, there is a one dimensional Coulomb branch,
parameterized by $\Phi =\phi +i\gamma$ which lives on a cylinder,
$\phi
\in \IR$ and $\gamma \in S^1$ 
of period $g^2$ (the radius of $S^1$ is of order $g$).  For $N_f>0$
there is also a $(2N_f-1)$-dimensional Higgs branch, which can be
parameterized by the gauge invariant operators $M^i_{\tilde j} = Q^i
\tilde{Q}_{\tilde j}$ 
subject to the constraint that rank$(M)\leq 1$, i.e.  $M^i_{\tilde
j}M^k_{\tilde l} = M^k_{\tilde j} M^i_{\tilde l}$. Classically the
Higgs branch intersects the Coulomb branch at $\phi = 0$. Since there
are no instanton corrections in this case, one might expect the
classical picture to continue to hold also in the quantum
theory. However, as we will see, perturbative effects change the
topology of the moduli space, and new degrees of freedom will be
needed to describe the quantum moduli space.

\subsec{The quantum moduli space of $U(1)$ gauge theories}

For large $\phi$, it appears that the Coulomb branch can be
consistently parameterized by the vacuum expectation value of the
chiral superfield $V = e^{\Phi / g^2}$. However, the metric for
$\gamma$ receives quantum corrections.  This has already been seen in
\refs{\irdyn,\swtd} in $N=4$ theories. Since the circumference of the
circle which $\gamma$ lives on is $g$, the topology of the moduli
space can be changed in perturbation theory. To determine the behavior
of the quantum theory near the origin of moduli space, note that the
Higgs branch (which classically intersects the Coulomb branch at
$\phi=0$) is invariant under the $U(1)_J$ symmetry which corresponds
to rotations around the circle. Therefore, the radius of $\gamma$ must
vanish where the two branches meet.

\fig{$U(1)$ $N=2$ gauge theory in three dimensions with massless flavors;
the classical moduli space and the quantum corrected moduli space.}
{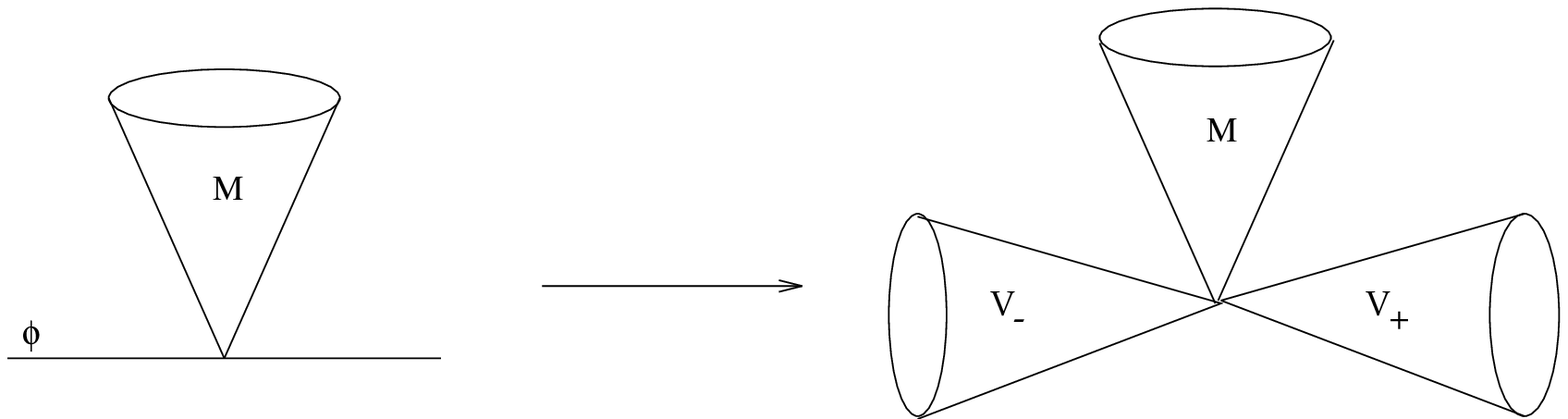}{12 truecm}
\figlabel\Onef

Thus, the Coulomb branch splits into two distinct regions,
and the moduli space looks (near the origin)
like an intersection of three cones (as in
fig. \Onef). The $\phi > 0$ half of the
Coulomb branch is parameterized by a field $V_+$, which is
semi-classically given by $V _+\sim e^{\Phi/g^2}$. Because the 
region of $\phi=0$ shrinks to a
point, in the quantum theory $V_+
\to 0$ there, and can take values in the entire complex plane
in this half of the Coulomb branch. Obviously, this means that a new
variable $V_-$, which semi-classically is given by $V_-\sim
e^{-\Phi/g^2}$, is needed to parameterize the second half of the
Coulomb branch.  So, our quantum picture of the Coulomb branch
involves two unconstrained chiral superfields $V_\pm$ to describe the
two distinct halves of the Coulomb branch.  The topology of the Higgs
branch is not changed by the quantum corrections, so we can still use
the mesons $M^i_{\tilde j}$ to parameterize this branch.

The metrics of the various branches in the semi-classical regions can
easily be computed. Far along the Higgs branch (large $\ev{M}$), the
K\"ahler potential of $M$ looks like $(M^\dagger M)^{1/2}$, while far
along the Coulomb branch (large $V_\pm$) the field $\Phi \sim \pm
\log(V_\pm)$ has a canonical kinetic term, and the Coulomb branch
looks like a cylinder.  However, at a distance of order $g$ {}from the
origin of moduli space, there will be strong corrections to these
metrics, as described above.

The quantum dynamics is constrained by the global symmetries, which
are
\eqn\nfgenui{
\matrix{  
\quad     & U(1)_R & U(1)_J& U(1)_A & SU(N_f) & SU(N_f)  \cr
          & & & &&\cr 
Q & 0 & 0 & 1 & {\bf N_f} & {\bf 1}\cr 
\widetilde Q & 0 & 0 & 1 & {\bf 1} & {\bf \overline N_f} \cr	
M & 0 & 0 &2  & {\bf N_f} & {\bf \overline N_f} \cr
V_\pm & N_f & \pm 1  & -N_f &{\bf 1} &\bf 1. \cr}} 
We chose the $U(1)_R$ charge of $Q$
and $\tilde{Q}$ to be zero, so that their fermions have charge $(-1)$.
The symmetry $U(1)_J$ corresponds to shifting the dual photon
$\gamma$; its current is $J_\mu = \epsilon_{\mu \nu \lambda} F^{\nu
\lambda}$.  It follows {}from our semi-classical identifications that
$V_\pm$ have charge $\pm 1$ under this symmetry.  The charges of
$V_\pm$ under the other global symmetries follow, as discussed in
\S2.4, {}from a one-loop diagram connecting the
currents to the gauge field.

For $N_f=1$ the three branches in fig. \Onef\ are all one complex
dimensional, and we propose that they are actually related near the
origin by a triality exchange symmetry.  At the origin, where the
three branches meet, there is a RG fixed point.  Another theory which
flows to the same IR fixed point can be described by the fields $M$,
$V_\pm$ with a superpotential
\eqn\wuinfi{W = -MV_+ V_-,}
which correctly gives the moduli space consisting of three cones,
parameterized by $V_\pm$ and $M$, which intersect at the point $V_\pm
=M=0$.  (The sign and normalization in \wuinfi\ are chosen for
convenience.)  The superpotential \wuinfi\ respects the global
symmetries \nfgenui.  Since \wuinfi\ is of degree 3, it is strongly
coupled in the IR and flows to a fixed point which we claim is the
same as that of $N_f=1$ SQED.

A (rather weak) check that $N_f=1$ SQED and the theory with $M$,
$V_\pm$ in \wuinfi\ flow to the same IR fixed point is that the parity
anomaly matching conditions discussed in the previous section are
satisfied.  In both the original $U(1)$ theory with $N_f=1$ and in the
theory with the fields $M$ and $V_\pm$, eqn.
\kabquant\ for the global $U(1)_R\times U(1)_J\times U(1)_A$ symmetries
gives $k_{RR}\in \IZ+\half$, $k_{JJ}\in \IZ$, $k_{AA}\in \IZ$,
$k_{RJ}\in \IZ$, $k_{RA}\in \IZ$, and $k_{AJ}\in \IZ$.

We can further test \wuinfi\ by giving the electron a complex mass
$m$, leading to $W = -V_+ V_- M + m M$. Integrating out $M$, we find
$V_+ V_- = m$; the two variables parameterizing the Coulomb
branch are related as expected for the free $N_f=0$ theory (where the
moduli space is just a cylinder).

In addition to this ``complex mass'' deformation, there are several
other deformations which cannot be described by a
superpotential. Consider, for example, adding a \FI\ term.  In the
original theory, this lifts the Coulomb branch, and we are left only
with the Higgs branch. As discussed in \S2.3, we can view this term as
a background $U(1)_J$ vector field. Thus, we can turn it on also in
the dual theory \wuinfi, where (using \nfgenui) it corresponds to a
real mass term for $V_+$ and $V_-$, which has the same effect on the
moduli space.

Another possibility is to add ``real masses'' for $Q$ and ${\tilde
Q}$. Opposite ``real masses'' for $Q$ and $\tilde Q$ may be absorbed
in the definition of the origin of $\phi$.  On the other hand, equal
masses $\tilde m$ (which breaks CP), corresponding to a background
$U(1)_A$ vector field, are physically significant.  On the Coulomb
branch, the ``effective real mass'' of $Q$ is now $(\phi+{\tilde m})$,
and that of $\tilde Q$ is $(-\phi+{\tilde m})$. There is no Higgs
branch since $Q$ and $\tilde Q$ are massless at different points on
the Coulomb branch, but classically the Coulomb branch remains
unlifted. However, if we integrate out the chiral multiplets for
$-{\tilde m} < \phi < {\tilde m}$ (where the two ``effective real
masses'' have the same sign), we will generate \redlich\ a
Chern-Simons term for the gauge field (with coefficient one), that
will give the gauge field a mass and lift this part of the Coulomb
branch. On the other hand, if we integrate out the chiral multiplets
for $\phi > \tilde m$ or for $\phi < -{\tilde m}$, the same diagrams
generate a \FI\ term for the gauge field (as discussed in \S2.4),
proportional to $\tilde m$. Thus, these regions of the Coulomb branch
are also lifted, and at most we can remain with discrete vacua near
$\phi = \pm {\tilde m}$. We can perform the same analysis also in the
dual theory \wuinfi. Here, a background $U(1)_A$ vector field
corresponds to giving a real mass to $M, V_+$ and $V_-$, so only a
single vacuum remains at the origin of moduli space, in agreement with
the analysis of the original theory. A similar analysis may be
performed for combinations of real mass and \FI\ terms.

For $N_f>1$, a similar analysis using the symmetries \nfgenui\ gives
(in a convenient normalization) 
\eqn\wuinfa{W = -N_f (V_+ V_- \det M)^{1/N_f}.}
This is reminiscent of the superpotentials found for $d=4$ $N=1$ SQCD
theories with $N_f > N_c+1$ \nati.
This superpotential describes the moduli space correctly away {}from
the origin and behaves in the expected way when we add (complex)
masses for some of the quarks.  However, \wuinfa\ is singular at the
origin of the moduli space, indicating the presence of new degrees of
freedom there. This is also indicated by the fact that the fields
$V_\pm$ and $M$ do not satisfy the parity anomaly matching conditions
for even values of $N_f$. We expect, again, that there is an
interacting fixed point at the origin.  A non-singular description of
that fixed point would have to include additional fields. In the next
section, we
propose another theory which flows to the same fixed point.

Again, we can add ``real mass'' terms for the chiral multiplets.  In
principle, we can add independent real masses for all the fields $Q^i$
and $\tilde{Q}_{\tilde i}$.  However, as discussed above, if we do not
give real masses of equal magnitude and opposite signs to $Q^i$ and
$\tilde{Q}_{\tilde i}$, CP is broken, and \FI\ or Chern-Simons terms
are generated at one-loop when integrating out the fermions.  To avoid
this, we discuss only the case where $Q^i$ ($\tilde{Q} _{\tilde i}$)
has real mass $\tilde m_i$ ($-\tilde m_i$).  Only the $N_f-1$ relative
mass terms are physically significant, as the average real mass can be
absorbed by a shift in $\phi$. For simplicity, consider the theory
with $N_f$ different real masses $\tilde{m}_i$. Classically, there are
$N_f$ one dimensional Higgs branches, parameterized by the diagonal
elements of $M_{\tilde i}^i$, which intersect the Coulomb branch
(parameterized by $\Phi$) at $\phi=\tilde{m}_i$.

\fig{$U(1)$ $N=2$ gauge theory in three dimensions with $N_f=4$ flavors
of different real masses: the quantum moduli space.}
{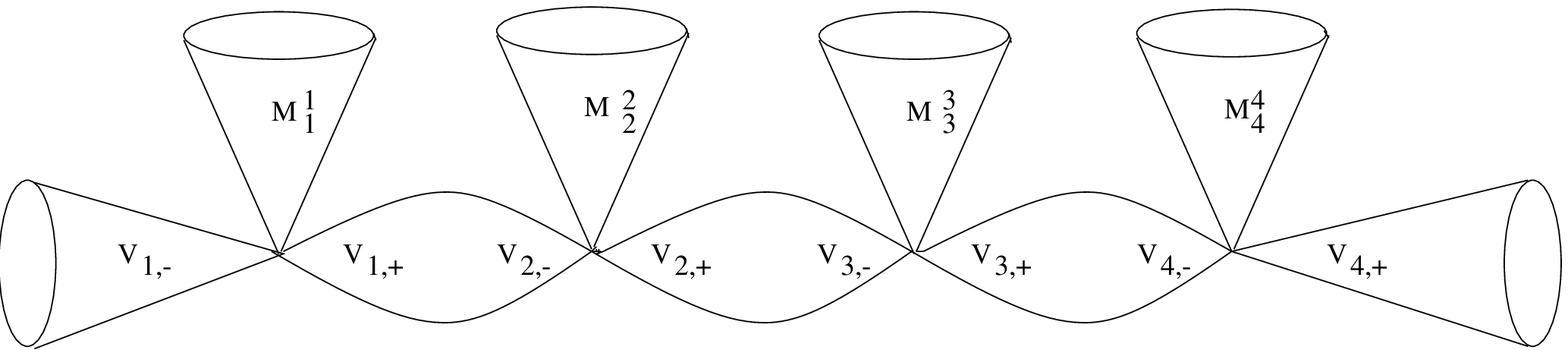}{15 truecm}
\figlabel\Onenf

At each such intersection point we have a $U(1)$ theory with $N_f=1$,
so, as discussed above, the Coulomb branch splits in the quantum
theory, and the complete moduli space looks like figure \Onenf. As
before, near each intersection point there are fields $V_{i,\pm}$,
which semi-classically behave as $e^{\pm (\Phi-\tilde{m}_i)/g^2}$,
with a superpotential of the form $-\sum _{i=1}^{N_f}
M^i_{i}V_{i,+}V_{i,-}$. This description involves two fields in each
of the middle $(N_f-1)$ regions of the Coulomb branch, while we expect
only one -- the semi-classical identifications suggest that we should
identify $V_{i,+} V_{i+1,-} = 1$ (up to a constant), which is
consistent also with the global symmetries.  The theory with $N_f$
quarks with different real masses is then described by
\eqn\uonereal{W = -\sum_{i=1}^{N_f}  M_{i}^iV_{i,+}V_{i,-} + 
\sum_{i=1}^{N_f-1} \lambda_i (V_{i,+} V_{i+1,-} - 1),}
where the $\lambda_i$ are Lagrange multipliers. As discussed in \S2.3, 
the superpotential
cannot depend on the real masses $\tilde m_i$, 
but we expect the K\"ahler potential
to depend on the $\tilde m_i$ in such a way that we recover the 
previous description when the $\tilde m_i$ are zero (or equal). 
Indeed, starting with \uonereal\ and 
integrating out the fields $V_{i<N_f,+}$ and $V_{j>1,-}$ yields
a superpotential of the
form $W = -N_f (V_{1,-}V_{N_f,+} M_1^1 M_2^2
\cdots M_{N_f}^{N_f}) ^ {1/N_f}$; this agrees with \wuinfa\
except for the absence of the off-diagonal mesons. This may be
relevant for understanding the additional degrees of freedom needed at
the origin for $N_f>1$.

\newsec{Mirror symmetry for $U(1)$ gauge theories}

$N=4$ $U(1)$ gauge theory with $N_f>1$ charged hypermultiplets has a
RG fixed point with a ``mirror'' dual description in terms of an
$(U(1)^{N_f})/U(1)$ gauge theory with $N_f$ hypermultiplets $q_i$ of
charge $1$ under the $i$'th $U(1)$ and charge $(-1)$ under the
$(i+1)$'th $U(1)$ (in cyclic order, with the sum of the $N_f$ $U(1)$'s
ungauged) \ismir. The $N=4$ vector multiplet includes an $N=2$ vector
multiplet and an $N=2$ chiral multiplet $\Psi$. Giving a mass to
$\Psi$ breaks $N=4$ to $N=2$.  The low energy theory is then the $N=2$
SQED theory discussed in the previous section.  Duality is preserved
under the RG flow, so mapping the $N=4$ breaking mass term to the dual
theory gives a dual description of the low energy $N=2$ theory.
Because these $N=2$ duals came {}from the $N=4$ mirror symmetry of
\ismir, we might still refer to them as ``mirror symmetry.''  (It is a
misnomer, though:  Unlike the $N=4$ case, where there are
$SU(2)_{R_1}\times SU(2)_{R_2}$ global symmetries which provide an
invariant way to distinguish between the Higgs and Coulomb branches,
which are exchanged under mirror symmetry, in $N=2$ theories there is no
invariant distinction between the Higgs and Coulomb branches.)
In principle, all $N=4$ mirror symmetries can be turned into $N=2$
mirror symmetries in this way\foot{In the brane description of the
mirror symmetry \hw, the operation of adding a mass to the adjoint
chiral superfield corresponds to rotating one of the NS 5-branes by
ninety degrees in two planes (say, the 48 plane and the 59 plane in
the conventions of \hw), an operation which can be performed while
preserving $d=3$ $N=2$ supersymmetry
\bdl. Similar issues were recently discussed in \barbon. 
This leads to a brane configuration which is T-dual to the
configurations discussed in \egk, and $SL(2,\IZ)$ duality in this
configuration leads to mirror symmetry in $N=2$ theories, as recently
discussed also in \bhoy.}.

Because the chiral multiplet $\Psi$ is neutral, there are two ways to
give it a mass.  We can either add a term $m\Psi^2$ to the
superpotential, or pair it with another singlet field $S$ 
which also gets a mass by a superpotential term $m S
\Psi$.  At low energies, either choice will give the $N=2$ theories 
discussed in the previous section, and the latter is simpler.
In the superpotential $W=S\Psi$ the field $S$ acts as a dynamical
\FI\ term and is mapped by mirror symmetry 
\ismir\ to a dynamical mass term, $W=S\sum _{i=1}^{N_f}q_i \tilde{q}^i$.
Including the $N_f-1$ neutral chiral fields in the $U(1)^{N_f-1}$
vector multiplets and their superpotential couplings to the 
matter fields, the $N=2$ superpotential of the dual is now
\eqn\spotmir{W = \sum_{i=1}^{N_f} S_i q_i \tilde{q}^i,}
where the $q_i$ are charged under the $U(1)^{N_f-1}$ as discussed above.
This theory gives a dual description of the $N=2$ SQED with
$N_f$ flavors RG fixed point discussed in the previous section.
(A similar proposal appeared in \bhoy\ via a brane construction.)

Let us compare the Coulomb branch of the original theory with
the Higgs branch of the dual. In the previous section we saw that the
Coulomb branch splits into two distinct regions, parameterized by 
$V_\pm$. The Higgs branch of the mirror theory, 
solving the classical D-term and F-term equations, is similarly
given by two distinct regions, parameterized by the gauge invariant
operators $N_- = q_1 q_2 \cdots q_{N_f}$ and $N_+ = \tilde{q}^1
\tilde{q}^2
\cdots \tilde{q}^{N_f}$, which intersect at $N_- = N_+ = 0$. Thus, 
we are led to identify $N_\pm$ with $V_\pm$. 
In the mirror theory, the quantum splitting of the Coulomb branch
is visible already at the classical level.

The Higgs branch of the original SQED theory is obtained {}from the
Coulomb branch of the mirror along with the $N_f$ chiral singlets
$S_i$. In the original theory, the $2N_f-1$ dimensional Higgs branch
is parameterized by the mesons $M_{\tilde i}^j$ subject to the
classical constraints $M_{\tilde i}^j M_{\tilde k}^l = M_{\tilde k}^j
M_{\tilde i}^l$, and intersects the Coulomb branch at
$M_{\tilde i}^j=V_\pm =0$. In the mirror theory, this moduli space
arises {}from the $N_f$ singlets $S_i$ and the $N_f-1$ dual photons,
and classically it intersects the Higgs branch when all these
variables, as well as $N_\pm$ vanish. Thus, the two theories
reproduce the same moduli space.

In fact, we can now identify the operators $M_{\tilde i}^i$
parameterizing the Higgs branch in the original theory with the
operators parameterizing the mirror Coulomb branch. It follows {}from
the map between mass terms of $N=4$ SQED and \FI\ terms in the mirror
\ismir\ that the diagonal elements of 
$M^i_{\tilde i}$ of the $N=2$ SQED theory are mapped to the $S_i$ of
the dual, for $i=1\dots N_f$.  The map between the remaining
components of $M_{\tilde i}^j$, subject to the classical constraint,
and the Coulomb branch moduli of the $U(1)^{N_f-1}$ dual is a bit more
difficult to obtain.
As in the previous section, we identify the two fields $W_{i, \pm}$
corresponding to the photon of the $i$'th $U(1)$.  These fields are
mapped under the duality as $W_{i,+}=M_{i}^{i-1}$ and
$W_{i,-}=M_{i-1}^i$.  The $i^{th}$ $U(1)$ has two charged fields with
masses $S_{i-1}$ and $S_i$.  Adding these mass terms to the analog of
\wuinfa\ for the $i^{th}$ $U(1)$ and integrating out the massive
matter yields the quantum relation $W_{i,+}W_{i,-}= S_i S_{i-1}$; this
corresponds to a classical constraint in the original theory.  The
remaining mesons $M_{\tilde{i}}^j$ can be identified using the
permutation symmetry of the electrons, which acts in the dual theory
via linear transformations on the $U(1)$ gauge fields.  The remaining
classical constraints on the mesons then follow {}from semi-classical
relations involving products of $W_{i,\pm}$ and $W_{j,\pm}$.

This dual description can be connected with our results of the
previous section.  For example, consider the $N_f=2$ case and flow to
$N_f=1$ by adding an operator $m M^2_2$ to the original theory.  The
mirror is then an $N=2$ $U(1)$ with a superpotential $W = S_1 q_1
\tilde{q}^1 + S_2 q_2
\tilde{q}^2 + m S_2$. The equation $\partial W/\partial S_2=0$
causes the gauge symmetry to
be broken, leaving three one-dimensional branches, parameterized
by the gauge invariant operators $S_1, q_1 q_2$ and $\tilde{q}^1
\tilde{q}^2$. These were identified above with $M^1_1, V_+$ and $V_-$
in the original theory, and the superpotential in terms of these
variables reproduces our result \wuinfi.

\newsec{The Vortex Interpretation}

$N=4$ and $N=2$ supersymmetric theories have semi-classical vortex
solutions which can be BPS saturated.  An understanding of these
states provides additional insight into the physics of mirror symmetry
in these theories.

Consider $N=2$ theories of $U(1)$ with $N_f$ flavors, or their mirrors
with products of $U(1)$ gauge symmetries.  In the presence of a real
\FI\ term, there are special vacua where only one 
of the expectation values for the $Q^i,\tilde Q_{\tilde i}$ is
non-vanishing; thus all holomorphic invariants of charged fields are
zero.  In such a vacuum, there are Nielsen-Olesen vortex solutions
\nielol\ in which the non-vanishing expectation value 
winds at spatial infinity, as does the gauge field:
\eqn\winding{
Q\sim \sqrt{\zeta}e^{\pm i\theta} \ ; A_\theta \sim \pm{1\over r} \ .  }
The covariant derivative $D_\theta Q$ falls off faster than $1/r$ and
so does not contribute a logarithmic divergence to the mass of the
vortex solution.  The BPS charge of this state is given by the winding
number $\pm 1$ of the gauge field times $\zeta$, and in this vacuum the
BPS bound is satisfied.\foot{As usual, there are solutions
with any value of the winding number, but the multi-winding vortices
are neutrally stable with respect to states with multiple vortices
with winding number one.  } The existence of a BPS charge is directly
tied \bpsbound\ to the existence of a Bogomolny bound \bogo.

However, other vacua on the Higgs branch do not have BPS vortices.  In
these more general vacua the vortex solutions still exist but do not
saturate the BPS bound.  This is possible because the vortices are not
in small multiplets.

In $N=4$ theories, there is an $SU(2)_R$ triplet of \FI\ terms which,
in terms of $N=2$, may be taken as a real parameter $\zeta_r$ and a
complex parameter $\zeta_c$.  Again vortices satisfy the BPS bound
only at special points on the Higgs branch.\foot{These statements are
related by dimensional reduction to similar statements about
instantons in two dimensions
\twodinst\ and strings in four dimensions \fourdstrng\ in
Abelian gauge theories.}

The BPS charge is easily seen to be proportional to the $U(1)_J$ shift
charge:
\eqn\qbps
{ Z = \zeta\int r d\theta A_\theta = \zeta\int d^2x \epsilon^{0\mu\nu}
F_{\mu\nu} = \zeta\int d^2x \ j^0 \ . }  This means that on the Higgs
branch there are massive vortex states of non-zero $J$ charge.  Since
the only holomorphic operators carrying this charge are $V_{\pm}\sim
e^{\pm\Phi}$, we may expect that the vortices are related to them.  In
fact, if we take $\zeta$ to zero (so that the special vacuum where the
vortices are BPS saturated becomes the origin of the Coulomb branch)
we may then move smoothly onto the Coulomb branch, where $U(1)_J$ is
spontaneously broken by expectation values for $V_{\pm}$.  This leads
to the natural interpretation that the Coulomb branch is associated
with vortex condensation --- the Higgs branch of the vortex theory.

 The exchange of Higgs and Coulomb branches is characteristic of
$N=4$ mirror symmetry, which also implies that the masses of the
mirror states should be given by the \FI\ terms in the electron
theory.  This is true precisely for the BPS vortices, and it is
therefore natural to interpret mirror symmetry (and its $N=2$
analogue) as an exchange of electron and vortex descriptions.

Let us check this interpretation for $N=4$ $U(1)$ with $N_f=1$.  For
non-zero $\zeta$ this theory has a unique vacuum, in which the vortex
$v_+$ and anti-vortex $v_-$ are BPS saturated and have mass $\zeta$.
We may write an effective theory for these fields which, in terms of
$N=2$, includes kinetic terms, a real mass $\zeta_r$ for the
vortices, and a superpotential
\eqn\vortexefL
{W = \zeta_c v_+ v_-\ \ .}  Note that the $N=4$ multiplet consisting
of $v_\pm$, whose mass is $\sqrt{\zeta _r^2+|\zeta _c|^2}$ is BPS
saturated.  It is not BPS saturated in $N=2$.

To go {}from this to $N=2$ with one flavor, we add a singlet $S$ and
couple it to the field $\Psi$ through the superpotential $W = S\Psi$.
This has the effect of replacing the complex \FI\ term $\zeta_c$ with
$S.$ In addition the equation of motion for $\Psi$ sets $S=-Q\tilde Q
= -M$.  Identifying $V_{\pm}$ as $v_{\pm}$ and making these
substitutions in
\vortexefL, we recover our result \wuinfi.  Here, the vortices are only
BPS saturated at the origin of the Higgs branch, where $M=0$.  Indeed,
the result \wuinfi\ implies that for non-zero $M$ the vortex masses
are given semi-classically (in the vortex effective theory) by
$\sqrt{\zeta^2+|M|^2}$ and therefore must be larger than $\zeta$ (even
when quantum corrections are included.)

Next consider $N_f=2$ for $N=2$ supersymmetry; one can easily check
that all of the following results can be lifted to the $N=4$ case as
well.  The electron theory has mesons $M^i_{\tilde j}$ which form a
$(\bf {2,2})$ of the $SU(2)\times SU(2)$ flavor symmetry.  The
vortices are as described above, and correspond to the operators
$V_\pm$.  The new feature of this theory is that its mirror has a
photon and a corresponding dual scalar $\tilde\gamma$.\foot{We
emphasize that this photon is not the magnetic dual of the photon
which couples to electrons -- it is a physically distinct field.}  A
shift in $\tilde\gamma$ is a global symmetry $U(1)_{\tilde J}$, which
should correspond to a global symmetry in the electron theory.  {}From
\ismir\ and the previous section, the generator of this global
symmetry is $T_3$ in the diagonal $SU(2)$ subgroup of the $SU(2)\times
SU(2)$ flavor symmetry.  This generator rotates $M_1^2$ and $M_2^1$ in
opposite directions (leaving $M_i^i$ unchanged) and so we identify
these operators with $W_{\pm}\sim e^{\pm(\tilde\phi+
i\tilde\gamma)}$.\foot{ A state with a single electron, which has
divergent energy due to its electric field $F^{0r} = \partial_\theta
\gamma = q/r$, causes $\gamma$ to wind at infinity: $V_+\sim
e^{iq\theta}$.  The mirror of this statement is that a charged mirror
field causes the phase of $W_+ = M_1^2$ to wind at infinity.  This is
precisely the winding one would expect for a global semi-classical
vortex, which indicates a connection between global vortices and the
charged fields in the mirror theory, a linkage already suggested by
the brane picture of \fourdstrng\ and \hw.  }  We can then see that
the $SU(2)\times SU(2)$ flavor symmetry, which is not visible
classically in the mirror theory, will appear in the far infrared in
the following way: the fields $M_1^1$ and $M_2^2$ (of dimension
$\half$ in the classical mirror theory) will combine in the far
infrared with the (classically dimensionless) fields $W_{\pm}$ to form
a $({\bf 2,2})$ representation.  This type of realization of a hidden
symmetry is analogous to many examples known in other dimensions.

We may also analyze the semi-classical vortices of the mirror theory
to recover states containing the original electrons.  The analogous
arguments lead one to identify $V_{\pm}$ with bilinears of the mirror
theory. Also, they show that the mesons $M^i_{\tilde j}$ represent BPS
saturated states at any point on the Coulomb branch at which $Q^i$ is
massive but $\tilde Q_{\tilde j}$ is not.  One can think of
$M^i_{\tilde j}$ as an electron $Q^i$ of bare mass $\tilde m_i$ which
is dressed by a massless positron $\tilde Q_{\tilde j}$ to eliminate
its logarithmic divergence.

As we mentioned earlier, these statements can easily be lifted back to
the $N=4$ theory, in which the hidden $SU(2)$ flavor symmetry of the
mirror is realized in the far infrared by the triplet $V_+,\Psi,V_-$.

For higher $N_f$ in either $N=2$ or $N=4$, the discussion is very
similar.  One can identify all of the scalars dual to the photons of
the mirror theory as phases of off-diagonal mesons $M_i^{i+1}$.  The
identification of the BPS vortex $V_{-}$ with $q_1q_2\cdots q_{N_f}$
can again be made by considering the mirror gauge theory with a real
mass $\zeta$; $q_1q_2\cdots q_{N_f}$ creates a state which is BPS
saturated at the points on the Coulomb branch where there is only one
massive field $q_i$, which when dressed with all the other $q_j$ forms
a state of finite energy $\zeta$.

For any $N_f>1$, one may add real masses to the original theory, along
with a \FI\ term.  For unequal masses this leaves the theory with
$N_f$ Higgs branches, each with a special vacuum where a vortex and
corresponding anti-vortex are BPS saturated.  These vortices can be
associated to the $2N_f$ operators $V_{i,\pm}$.  In the mirror, all
gauge symmetries are broken, and the fields $q_i,\tilde q^{\tilde j}$
are gauge singlets which can be related directly to the $V_{i,\pm}$.

In summary, the mirror symmetries exchange Nielsen-Olesen vortices
with logarithmically confined states of mirror electrons; similarly
the confined states of the original theory appear as Nielsen-Olesen
vortices of the mirror.  The exchange of Higgs and Coulomb branches,
masses and \FI\ terms, {\it etc.}, is natural in this language.
Furthermore, in both the $N=4$ and $N=2$ $U(1)$ theories, the
existence of hidden symmetries \ismir\ realized only in the far
infrared predicts the presence of certain operators with certain
charges, which we have identified as the chiral operators $V_{\pm},
W_{i,\pm}$.  Here, we have understood these operators as associated
with semi-classical Nielsen-Olesen vortices.\foot{Further checks on
this picture can be provided using the brane construction of \hw, and
one finds general consistency.  However, interpretation of this
construction is subtle, as no sign of the quantum corrections
associated with logarithmic confinement is to be found at the level of
classical branes.}

\newsec{$SU(2)$ gauge theory}

For $SU(2)$ gauge theories the Coulomb branch is parameterized by the
expectation value of $\Phi$, whose scalar component is $\phi +i\gamma$
with $\phi\in R^+$ and $\gamma \simeq \gamma + g^2$ (at tree level).
As discussed in sect. 2.5, the natural coordinate on the Coulomb branch is
$Y$, given semi-classically by $e^{\Phi /g^2}$.

To avoid having to include a Chern-Simons term, we will discuss
theories with an even number, $2N_f$, of chiral multiplets $Q^i$ in
the ${\bf 2}$ representation.  For $N_f>0$ the classical moduli space
also has a $4N_f-3$ complex dimensional Higgs branch which attaches to
the Coulomb branch at $\phi=0$.  As in 4d, the Higgs branch can be
labeled by the mesons $M_{fg}=Q^c_fQ^d_g\epsilon _{cd}$,
$f,g=1,\dots,2N_f$.  For $N_f\geq 2$ the $M_{fg}$ are classically
constrained by rank$(M)\leq 2$, i.e.  $\epsilon^{i_1i_2\cdots
i_{2N_f}} M_{i_1i_2} M_{i_3i_4} = 0$ (which for $N_f=2$ is just $\pf
M=0$).

The quantum corrections to the classical moduli space are constrained
by the global flavor symmetries, which are:
\eqn\nfiigc{
\matrix{  
\quad     & U(1)_R & U(1)_A & SU(2N_f)_F  \cr
          & & & \cr Q & 0 & 1 & {\bf 2N_f} \cr M & 0 & 2 & {\bf
	N_f(2N_f-1)} \cr Y & 2N_f-2 & -2N_f & {\bf 1}. \cr }} 
The charges assigned to $Y$ follow {}from one-loop perturbative
effects, similar to those described in \S2.4. They are consistent with
the fermion zero modes of an
instanton on the Coulomb branch, which is weighed by $Y^{-1}$: there
are two gluino zero modes and $2N_f$ quark zero modes, $\psi
_{i=1\dots 2N_f}$.

\subsec{Pure Yang-Mills -- no supersymmetric vacuum}

This is the theory analyzed in \ahw. The instanton has two zero modes,
which is the correct number to generate a superpotential, and indeed 
the instantons generate a superpotential \ahw\
\eqn\wahw{W = {1\over Y}.}
(We do not explicitly write an overall scale needed on dimensional
grounds.)  This superpotential is exact, as it is the unique result
which respects the global $U(1)_R$ symmetry \nfiigc.  The classical
degeneracy of the Coulomb branch is lifted by \wahw\ and the theory
has no vacuum.  This is reminiscent of the $d=4$ $SU(2)$ gauge theory
with $N_f=1$, where there is no Coulomb branch but a similar
superpotential is generated on the Higgs branch \ads.

\subsec{$N_f=1$ : quantum merging of Higgs and Coulomb branches}

When the quark flavor is massless, there is classically a one complex
dimensional Higgs branch, labeled by the expectation value of the
meson $M=Q_1Q_2$, which intersects the one complex dimensional Coulomb
branch at the origin, $M=\phi=0$.  It is impossible to describe this
in terms of $M$ and $Y$ in a holomorphic way.

Because all fields in \nfiigc\ are neutral under $U(1)_R$, it is
impossible to form a superpotential; therefore $W=0$ and there is an
exactly degenerate quantum moduli space of vacua.

The quantum moduli space of vacua, however, differs {}from the classical
moduli space because of quantum effects associated with instantons.
The quantum moduli space of vacua is given by the fields $M$ and $Y$
subject to the constraint
\eqn\suiiqc{MY=1.}
With this
constraint, \wahw\ is reproduced at low energy upon adding $W=m_QM$
(and matching the scales in an appropriate way). 
The space \suiiqc\ is a smooth quantum deformation of the singular
classical moduli space.  This is analogous to what happens in $d=4$,
$N=1$ supersymmetric $SU(2)$ theories with $N_f=2$ flavors \nati.

To summarize, classically there are two distinct branches, Higgs and
Coulomb, which intersect at the origin.  Quantum-mechanically, they
smoothly merge near the origin to form a single branch of
moduli space subject to the constraint \suiiqc. The results derived
here for this theory seem different {}from the solution proposed in 
\gomez.

\subsec{$N_f \geq 2$ : moduli space with non-trivial RG fixed
points at the origin}

For $N_f=2$, there is a unique superpotential consistent with
holomorphy and the symmetries with the charges \nfiigc:
\eqn\suiic{W=-Y\pf M.}
This reproduces the quantum constraint \suiiqc\ upon adding a mass
term $W=mM_{34}$ and integrating out the massive fields.  The theory
without mass terms has a quantum moduli space of vacua given by
expectation values of $M_{fg}$ and $Y$ subject to constraints coming
{}from the equations of motion of \suiic:
\eqn\mscon{\pf M=0, \qquad YM_{fg}=0.}
This is analogous to the situation for the $d=4$ $SU(2)$ theory with
$N_f=3$ \nati.  There are thus distinct Higgs and Coulomb branches,
with $M\neq 0$ and $Y=0$ on the Higgs branch and $Y\neq 0$, $M=0$ on
the Coulomb branch.  This is similar to the classical moduli space,
though now the branches touch at the origin, $M=Y=0$, rather than at
$M=\phi=0$. 
Because \suiic\ is of degree three in the fields $Y$ and $M_{fg}$, as
discussed in sect. 2.1, this theory flows to an interacting RG fixed
point.  We argue that the original $SU(2)$ theory with $N_f=2$ flavors
flows to the same fixed point.

As a (rather weak) check that the original $SU(2)$ with
$N_f=2$ theory and the theory with fields $M_{fg}$ and $Y$ with
superpotential \suiic\ flow to the same fixed point, we note that
their parity anomalies match.  For the $U(1)_R\times U(1)_A$ part, 
in both theories we have $k_{RR}\in \IZ+\half$, $k_{AA}\in \IZ$, and
$k_{RA}\in \IZ$.  For the $SU(4)$, in both the original theory,
which has two fields in the ${\bf 4}$, and the dual, which has the
field $M$ in the ${\bf 6}$, there is no parity anomaly, $k_{SU(4)}\in
\IZ$.

For $N_f > 2$, the symmetries \nfiigc\ determine the superpotential to be
(with a convenient normalization)
\eqn\weffi{W=-(N_f-1)\left(Y
\pf M \right) ^{1/(N_f-1)}.}
The branch cut singularity is analogous to the effective
superpotentials found in $d=4$, $N=1$ SUSY for $N_f\geq 4$.  As in the
$d=4$ case \nati, we interpret the singularity at the origin in
\weffi\ as a signature of important new degrees of freedom there.
We expect that there is an interacting RG fixed point at the origin.

\subsec{Adding real mass terms}

Next, we consider adding real mass terms for the quarks.
Unlike the U(1) case, we cannot absorb any such mass terms by
shifts of the scalar field $\phi$ in the vector multiplet.  
To preserve CP, whenever we give one doublet a real mass $\tilde{m}$
we should give another doublet a real mass $(-\tilde{m})$. Otherwise,
Chern-Simons and/or \FI\ terms will be generated along the Coulomb
branch, as we discussed in the $U(1)$ case. 
For simplicity, we only discuss the case of equal masses $\tilde{m}$
for $N_f$ doublets, and masses $(-\tilde{m})$ for the other $N_f$
doublets. Other cases may be obtained by a combination of the
discussion here and the discussion of the previous section on $U(1)$
theories with different real masses.

These real mass terms break the global flavor symmetry {}from
$SU(2N_f) \times U(1)_A$ to $SU(N_f) \times SU(N_f) \times U(1)_A
\times U(1)_B$ (they correspond to a background vector field in
a $U(1)_B$ subgroup of $SU(2N_f)$). The mesons which can obtain
expectations values for $\tilde{m}>0$ are now given by
$M^i_{\tilde j} = Q^i \tilde{Q}_{\tilde j}$, where the $Q^i$ are the
doublets with positive real mass and the $\tilde{Q}_{\tilde j}$ are
the doublets with negative real mass term. The Higgs branch is now
$2N_f-1$ dimensional, and classically it intersects the Coulomb branch
at $\phi =
\tilde{m}$.  

The region near $\phi = \tilde{m}$ looks like a $U(1)$
theory with $N_f$ massless flavors.  Our previous analysis thus shows
that the Coulomb branch splits into two regions, parameterized by
$V_\pm$ with, at the perturbative level, a superpotential of the
form $W = -N_f (V_+ V_-
\det M)^{1/N_f}$.
In the semi-classical regimes, $V_+
\sim e^{(\Phi-\tilde{m})/g^2}$ (for $\phi \gg
\tilde{m} \gg 0$) and $V_-\sim e^{(\tilde{m}-\Phi)/g^2}$ (for
$\tilde{m} \gg \phi \gg 0$).  

There can be additional instanton contributions 
to the superpotential.
Semi-classically, for quarks of real mass $\tilde{m}$, there are zero
modes in the background of the instanton if and only if $\phi >
|\tilde{m}|$
\callias. Thus, we expect that in the region $\phi < |\tilde{m} |$ an
instanton will have only the two gluino zero modes and can contribute
to the superpotential. For $\phi > |\tilde{m}|$ because of the
additional quark zero modes, there will be no instanton contribution
to the superpotential.  Without incorporating the splitting of the
Coulomb branch described in fig. 1, these statements seem paradoxical
(since the superpotential is holomorphic).  Taking the splitting into
account, the instanton contribution to the superpotential is simply
given by an additional term $W_{inst}=V_-$, which is indeed non-zero
only in the region that classically corresponded to $\phi <
\tilde{m}$. Combining with the perturbative superpotential written above,
yields the superpotential
\eqn\surmpot{W = -N_f (V_+ V_- \det M )^{1/N_f} + V_-.}
(Again, the normalization is chosen for convenience.)  Upon taking
into account the one-loop corrections to the global charges of the
$V_{\pm}$ fields, discussed in \S2.4, \surmpot\ is consistent
with the global symmetries.

Consider, for example, the case of $N_f=1$. Then \surmpot\ is $W = V_-
(1 - V_+ M)$, which is similar to our previous description of this
case (eqn. \suiiqc) except that $V_-$ has become a dynamical field
instead of a Lagrange multiplier.  The moduli space is the same as in
the massless case, and we expect the mass of $V_-$ to depend on
$\tilde{m}$ such that it is massless as $\tilde m\rightarrow
\infty$.   Adding a
complex mass $W = m M$ and integrating out the quarks, we find a
constraint $V_+ V_- = m$ and a superpotential $W = V_-
\sim 1/V_+$, both as expected for the low energy theory with
$N_f=0$. For higher $N_f$, we can no longer smoothly connect
\surmpot\ to the Lagrangian with no real masses, since some degrees of
freedom (corresponding to mesons of the form $Q^i Q^j$ and
$\tilde{Q}_{\tilde i} \tilde{Q}_{\tilde j}$) become massless only when
$\tilde{m}=0$, and do not appear in our Lagrangian for the massive
case. The qualitative behavior of the theory is, however, still the
same as in the massless case. There is a one dimensional Coulomb
branch parameterized by $Y = V_+$ which can take any value, which we
identify with the semi-classical region of $\phi >
\tilde{m}$, and a Higgs branch which intersects it at $Y = 0$.

\newsec{$SU(2)$ theories {}from four to three dimensions}

It is interesting to interpolate between the above results and those
of $d=4$ $N=1$ theories \refs{\ads, \nati} by considering these
theories on a circle of radius $R$. The scalars in the vector
multiplet now live on a circle of radius $1/R$, and the $d=4$ and
$d=3$ couplings are related classically by $1/g_3^2 = R/g_4^2$. In
this section we discuss the superpotential for finite values of
$R$. We can relate this to the $d=4$ results by taking $R \to
\infty$. In order to keep $g_4$ fixed, this means we need to take $g_3
\to 0$, and both circles corresponding to the Coulomb branch shrink to
zero size in this limit.  Thus, we should integrate out the fields
corresponding to the Coulomb branch in order to get the $d=4$
results. For any radius $R > 0$, the $d=4$ instantons break the
$U(1)_R$ symmetry and, for $N_f>0$, also the $U(1)_A$ symmetry -- but
a linear combination of the two is preserved.

\subsec{$SU(2)$ with $N_f=0$}

This case appeared in \swtd, where it was connected with the
results for $N=4$ theories by breaking to $N=2$.  The leading term
associated with finite radius was found to modify \wahw\ as
\eqn\wahwr{W={1\over Y} + \eta Y,}
where $\eta \sim e^{-1/Rg_3^2}$ is the four dimensional
instanton action, which in the four dimensional limit becomes
$\eta \sim e^{-1/g_4^2} \sim \Lambda_4^{3N_c-N_f}$ (in our case this
is simply $\Lambda_4^6$). 
However, we can also understand the second term
directly, as arising {}from $d=3$ 
instantons which are related by a large gauge transformation around 
the compact circle to the usual instantons\foot{This is obvious in the
brane construction of these theories.} \leeyi. $d=4$ instantons
have too many zero modes in this case so they do not contribute to the
superpotential.
For finite $R$, the superpotential \wahwr\
has the two vacua expected in the $d=4$ theory. In the $R \to \infty$
limit, we can integrate out $Y$ and get $W \sim \Lambda_4^3$,
corresponding to $d=4$ gaugino condensation. In the 3d limit $R=\eta
=0$, the two vacua run off to $Y\rightarrow \infty$, and the theory has
no vacuum.

\subsec{$SU(2)$ with $N_f>0$}

In all these theories, the $d=3$ instanton with twisted boundary
conditions around the circle still has no quark zero modes, so it will
always contribute to the superpotential a term $W_\eta = \eta Y$, which
will lift the Coulomb branch. Thus, for $N_f=1$,
the theory is described by the superpotential
\eqn\suiiqcr{W = \lambda(MY - 1) + \eta Y,}
where $\lambda$ is a Lagrange multiplier to impose the constraint
\suiiqc. For $N_f=1$ only,
the $d=4$ instantons can also 
contribute to the superpotential (if two gluino zero modes and two
quark zero modes are lifted together), and the global symmetries force
their contribution to be of the
form $W = \eta / M$, but this is identified with the second term in
\suiiqcr\ 
using the constraint so we do not have to take it into account
separately.  Upon integrating out $Y$ by imposing the constraint,
\suiiqcr\ leads to $W\sim \eta /M$, as in the four
dimensional theory
\ads. The theory has no supersymmetric vacua for any $R > 0$.

As for $N_f=0$,
we can find this result also by starting with the curve describing the
$N=4$ theory and giving a mass to the adjoint chiral multiplet \swtd.
For $N_f=1$, in order to repeat the computation of \swtd, we start
with the superpotential 
\eqn\nfonepot{W = \lambda(y^2 - x^3 + x^2 u -
(\Lambda_4^{N=2})^6) + \epsilon u,} 
which incorporates the curve as a constraint, and
includes a mass $\epsilon$ for the adjoint field $u$. Integrating out
$\lambda, u$ and $y$ as
in \swtd, we find
\eqn\threednfone{W = \epsilon ((\Lambda_4^{N=2})^3 x + 1 / x^2).}
Now, identifying $\epsilon/x$ with $e^{-\Phi/g^2}$ as in \swtd, and
relating the $d=4$ scales in the usual way by $(\Lambda_4^{N=1})^5 =
(\Lambda_4^{N=2})^3 \epsilon^2$,
we find the three dimensional superpotential to be
\eqn\morew{W = e^{-1/Rg^2+\Phi/g^2} + {1\over \epsilon} e^{-2\Phi/g^2},} 
and the
last term vanishes in the limit $\epsilon \to \infty$ leaving us with
the second term of \suiiqcr. The same result may be obtained also for
higher values of $N_f$. By using only the Coulomb branch
description of the $N=4$ theory we cannot obtain in a simple way the
correct dependence on the meson fields.

For $N_f=2$, the theory is described by the superpotential
\eqn\suiicr{W=-Y\pf M+\eta Y.}
Now, upon integrating out $Y$, \suiicr\ gives the quantum-modified
moduli space constraint of \nati: $\pf M\sim \eta$. Note that again we
find that the $d=4$ instanton result is reproduced by $d=3$ instantons.
For finite $R$
this constraint is implemented as the equation of motion of $Y$, and
$Y=0$ in the vacuum.

For $N_f>2$, adding $W_\eta =\eta Y$ to \weffi\ and integrating out
$Y$ leads to
\eqn\weffir{W \sim (\eta ^{-1}\pf M )^{1/(N_f-2)},}
which is the correct effective potential for the 4d theories
\nati. Again, the moduli space for finite $R$ is the same as in the
$d=4$ theory, with $Y=0$ at the vacuum.

\subsec{$SU(2)$ with real masses on a circle}

Real masses are also possible in the four dimensional theory
compactified on a circle. In this case they should be thought of as
Wilson loops of background vector fields, instead of VEVs of their
scalar components, so the masses also naturally live on a circle of
radius $1/R$.

Let us begin with the $N_f=1$ case, with a real mass $\tilde{m}$ of
opposite sign for the two doublets. Semi-classically, we can take $0
\leq \phi,\tilde{m} \leq 1/2R$. The standard instanton (with action
$e^{-\Phi/g^2}$)
has quark zero modes if and only if $\phi > \tilde{m}$, while the
twisted boundary condition instanton (with action
$e^{-(1/R-\Phi)/g^2}$) has quark zero modes if and only if $\phi <
\tilde{m}$. Using our previous description for the $N_f=1$ case with a
real mass, we obtain that the superpotential on a circle is
\eqn\spotcirc{W = -V_- V_+ M + V_- + \eta V_+.}
As before, there are no supersymmetric vacua for any value of $R$.

For $N_f > 1$ with equal real masses, we similarly obtain
\eqn\spotcircn{W = -N_f (V_- V_+ \det M)^{1/N_f} + V_- + \eta V_+.}
The Higgs branch now remains unlifted, with a constraint on $\det M$
for $N_f=2$ and no constraints for $N_f > 2$. The vacuum obeys $V_- =
(\det M \eta^{-1})^{1/(N_f-2)} = \eta V_+$, so the Coulomb branch is
lifted.  For $R \to \infty$ we can integrate out $V_-$ and $V_+$, and
get the correct $d=4$ limit except for the absence of some of the
mesons, which became massive due to the Wilson loop in $d=3$ but
cannot be ignored in the $d=4$ limit.

If we allow different real masses, some of the Coulomb branch may
remain unlifted. For instance, for $N_f=2$ with two different real
masses, the superpotential describing the theory (derived by the same
methods as above) is
\eqn\spotcirctwo{W = -V_{1,-} V_{1,+} M_1^1 - V_{2,-} V_{2,+} M_2^2
+ \lambda(V_{1,+}
V_{2,-} - 1) + V_{1,-} + \eta V_{2,+},}
where $\lambda$ is a Lagrange multiplier (as in the $U(1)$ theory with
different real masses). In this case a stable vacuum exists, in which
$V_{1,-} = V_{2,+} = 0$ and $M_1^1 M_2^2 = \eta$ 
(similar to the constraint
we found before for $N_f=2$). Note that classically the Higgs branches
corresponding to $M_1^1$ and $M_2^2$ are disjoint (with the Coulomb
branch connecting them), but they merge together in the quantum
theory. Similar phenomena occur with more different real masses.

\newsec{$SU(N_c)$ gauge theories with $N_c > 2$}

Writing the adjoint scalar in the form $\phi =diag(\phi _1, \dots \phi
_{N_c})$, with $\sum _{j=1}^{N_c} \phi _j=0$, the wedge of the Coulomb
branch Weyl chamber can be taken to be $\phi _1\geq \phi _2\geq \dots
\geq \phi _{N_c}$. The instanton factors \monj\ can be taken to be
$Y_j^{-1}$ with, semi-classically,
\eqn\suninj{Y_j
\sim e^{(\Phi_j-\Phi_{j+1})/g^2}, \qquad j=1,\dots, N_c-1.}  

We consider the theories with $N_f$ flavors of ${\bf
N_c}+{\bf{\overline{N_c}}}$; with this choice we can avoid introducing
a Chern-Simons term.  For $N_f>1$, in addition to the Coulomb branch,
the theory has a Higgs branch with $SU(N_c)$ generically broken to
$SU(N_c-N_f)$ for $N_f<N_c-1$ and completely broken for $N_f\geq
N_c-1$.  As in 4d, the Higgs branch can be parameterized by the $N_f^2$
mesons $M^i_{\tilde j}$ for $N_f<N_c$ or by mesons along with baryons,
subject to classical constraints, for $N_f\geq N_c$.

Global symmetries, holomorphy and known limits, as usual \nonren,
powerfully constrain the dynamics.  The $j$-th instanton, which is
weighted by $Y_j^{-1}$, has two gaugino zero modes.  To begin with, we
consider the theory with massless flavors, setting the complex and
real masses to zero.  At a generic point on the Coulomb branch, using
the analysis of
\callias\ (easily generalized to $SU(N_c)$ since the instantons are
just embeddings of $SU(2)$ instantons), we find that the $j$-th
instanton has $2N_f\delta _{j,K}$ quark zero modes, where $K$ is the
value which satisfies $\phi _K > 0 > \phi _{K+1}$\foot{It is easy to
see the $K$ dependence of the quark zero modes in the brane
construction of these theories, which is related by a rotation of one
of the NS 5-branes to the construction of \hw. The $\phi_j$'s in this
construction are positions of the D3-branes, while the instantons are
Euclidean D-strings stretching between them, and they will have quark
zero modes for a particular quark (semi-classically) if and only if
the corresponding D-string intersects the D5-brane which gives rise to
this quark.}.  The global symmetries and charges of the fields
(including the one-loop corrections described in \S2.4) are thus
\eqn\nfgengc{
\matrix{  
\quad     & U(1)_R & U(1)_B& U(1)_A & SU(N_f) & SU(N_f)  \cr
          & & & &&\cr 
Q & 0 & 1 & 1 & {\bf N_f} & {\bf 1}\cr 
\widetilde Q & 0 & -1 & 1 & {\bf 1} & {\bf \overline N_f} \cr	
M & 0 & 0 &2  & {\bf N_f} & {\bf \overline N_f} \cr
Y_{j\neq K} & -2 & 0 &0 &{\bf 1} &\bf 1 \cr
Y_K & 2(N_f-1) & 0 & -2N_f & {\bf 1} & {\bf 1} \cr 
Y & 2(N_f-N_c+1) &0 & -2N_f & {\bf 1} &{\bf 1}. \cr }} 
Here $Y\equiv \prod _{j=1}^{N_c-1}Y_j\sim e^{(\Phi _1 -\Phi _{N_c})/g^2}$
is included for later convenience.

The dependence on $K$ in \nfgengc\ reflects the fact that the theory
with matter is defined in $N_c-1$ {\it sub-wedges} of the Weyl
chamber, corresponding to the choice of $K=1\dots N_c-1$.  The fact
that the description is not smooth at the $N_c-2$ boundaries of the
sub-wedges, $\phi _j=0$ (for $j=2,\dots,N_c-1$), is possible because
there are massless fields on these boundaries: this is where
components of the quarks classically become massless and where the
Higgs branch connects to the Coulomb branch.  Indeed, along these
boundaries of the sub-wedges the low energy theory is governed by a
particular $U(1)\subset SU(N_c)$ which has $N_f$ massless flavors.
Thus, in accord with our analysis in \S3, we expect the Coulomb branch
to split along this sub-locus, with different variables describing the
regions on either side of the boundary.  Altogether, for $N_f>0$ the
bulk of the Coulomb branch splits into $N_c-1$ regions, with
(generally) different variables describing the various regions.

\subsec{The pure SYM theory, $N_f=0$.}

For $N_f=0$ we do not have the splitting described above of the
Coulomb branch into sub-wedges.  The instanton contributions to the
superpotential along the coulomb branch are
\eqn\sunymw{W=\sum _{j=1}^{N_c-1} {1 \over Y_j}.}
Every term in \sunymw\ is generated by the ``fundamental'' instantons
in the same way as in the $SU(2)$ theory \ahw.  The superpotential
\sunymw\ is not the most general one consistent with the symmetries
\nfgengc. However, a simple analysis shows that in fact \sunymw\ is
exact. First, since there are no interacting massless degrees of
freedom on the Coulomb branch, $W$ must be single valued as each
$Y_i\rightarrow e^{2\pi i}Y_i$. This leaves the possibility of terms
like $Y_i/Y_j^2$.  However, for every one of these terms, we can find
a way to rescale all $\phi_i\rightarrow\infty$ consistent with
$\phi_1>\phi_2>\cdots>\phi_{N_c}$ such that $Y_i/Y_j^2$ diverges. This
leaves \sunymw\ as the only form which is compatible with the
symmetries, holomorphy and the asymptotic behavior.

The superpotential \sunymw\ was obtained in \kv\ by considering M
theory on Calabi-Yau fourfolds.   For the non-supersymmetric version,
the scalar potential corresponding to \sunymw\ was also obtained long
ago by \waddas, generalizing the $SU(2)$ analysis of \polyakov.

This theory has no stable supersymmetric vacuum.

\subsec{Theories with massless quarks}

It follows {}from \nfgengc\ that there can be instanton contributions to
the superpotential for any $N_f$ coming {}from the $N_c-2$ instantons
$Y_{j\neq K}$, $j=1\dots N_c-1$:
\eqn\coulombpot{W _{inst}= \sum_{j\neq K}{1\over Y_j},}
which will lift most of the Coulomb branch. Note that the variables
$Y_i$ in the different regions (corresponding to different values of
$K$) are generally not the same, due to the splitting effect described
above. Similar contributions due to ``fundamental'' instantons will
exist in any region of the moduli space where semi-classically there
are two eigenvalues of the same sign. Thus, we expect that at most the
one complex dimensional sub-locus of the Coulomb branch
in which classically
\eqn\remaining{\phi_1 > \phi_2 = \cdots = \phi_{N_c-1} = 0 >
\phi_{N_c} = -\phi_1}
may remain unlifted.

In order to have a full description of the moduli space, we need to
repeat our procedure in the previous sections. We should first find a
superpotential which correctly describes the quantum moduli space
before the non-perturbative corrections, and which will have
(generally) different variables in each of the $N_c-1$ regions of the
Coulomb branch. Then, we should add to it the instanton corrections of
the form \coulombpot. For the $SU(3)$ case, we will explicitly
perform these computations below. For higher $N_c$, the superpotentials
describing the quantum-corrected Coulomb branch are complicated, and
we have not written them down explicitly. 

There is always one instanton factor, corresponding to the sum of the
simple roots, $Y =
\prod_{i=1}^{N_c-1} Y_i \sim e^{(\Phi_1-\Phi_{N_c})/g^2}$, which can be
globally defined throughout the Coulomb branch.  Whenever some part of
the Coulomb branch (which will correspond to \remaining) remains
unlifted, it can be parameterized by $Y$. Thus, we can always
integrate out all the other fields appearing in the superpotential,
and remain with an effective description that includes only the field
$Y$ and the fields parameterizing the Higgs branch.

For $N_f < N_c-1$, \nfgengc\ completely determines the
superpotential for these fields to be (with a convenient normalization)
\eqn\spotsmall{W = (N_c-N_f-1) (Y \det M )^{1/(N_f-N_c+1)};}
this is similar to the $d=4$ superpotentials for $N_f < N_c$ \ads.
With \spotsmall, there is no stable vacuum, so our procedure of
integrating out only some of the massive fields is not really
justified. 

For $N_f = N_c-1$, we obtain a quantum constraint on the moduli space,
of the form $Y\det M =1$, generalizing \suiiqc.  This is similar to
the quantum constraints found for $N_f = N_c$ in the $d=4$ theories
\nati. The moduli space is again in a merged Higgs and Coulomb 
branch, as we found for $SU(2)$ with $N_f=1$, with the small $\det M$
region of the Higgs branch corresponding to large values of $Y$, far
out on the Coulomb branch.  In particular, the points with $\det M=0$,
where part of the gauge symmetry would classically be unbroken, are at
infinite distance on the Coulomb branch.

For $N_f \geq N_c$, the superpotentials can
involve also baryonic operators $B$ and $\tilde{B}$ in addition to
\spotsmall\ continued to $N_f\geq N_c$.
For $N_f=N_c$, we argue that, much as in \suiic, the theory is
described by the fields $Y$, $M^f_{\tilde g}$, $B$, and $\widetilde B$
with the superpotential
\eqn\spotmid{W = -Y (\det M - B \tilde{B}).} 
Because of the cubic $YB\tilde B$ term, this theory flows to a
non-trivial fixed point in the IR; the original theory flows to the
same fixed point, which seems to be the same fixed point as the one we
found for the $U(1)$ theory with $N_f=1$ \wuinfi.  One check of this
is that the parity anomaly matching conditions discussed in \S2.4 are
satisfied: it follows {}from \nfgengc\ and \kabquant\ and \nonabk\
that, in both the $N_f=N_c$ SQCD theory and the theory with fields
$M^j_{\tilde i}$, $B$, and $\tilde B$ with \spotmid, $k_{RR}\in
\IZ+\half (1+N_c^2)$, $k_{BB}\in \IZ$, $k_{AA}\in
\IZ$, $k_{SU(N_f)_L}\in \IZ +\half N_c$, $k_{SU(N_f)_R}\in \IZ +\half
N_c$.  As another check, \spotmid\ is compatible with reproducing the
correct Higgs branch.  In particular, upon Higgsing to $SU(2)$ with
$N_f=2$ it properly reproduces \suiic.  Also, \spotmid\ properly
yields \spotsmall\ upon adding mass terms and integrating out flavors.
Finally, as will be discussed below, \spotmid\ properly connects to
the 4d result of \nati.

We can easily extend this discussion to $U(N_c)$ gauge groups with $N_f
\le N_c$.  It is enough to study the case $N_f=N_c$; lower values of
$N_f$ can be obtained by integrating out some quarks.  Ignoring the
$U(1)$ factor, we have the superpotential \spotmid.  We can now gauge
the $U(1)$ factor.  In this description the $U(1)$ dynamics is that of
$U(1)$ with one flavor: $B$ and $\tilde B$.  This theory is described by
the gauge invariant combination $X= B\tilde B$, two more chiral fields
$V_\pm$ and a superpotential $-XV_+V_-$.  Adding this to \spotmid\ leads
to $W=-Y(\det M - X) - XV_+V_-$.  Integrating out the massive
fields $Y$ and $X$ we find
\eqn\wunc{W=- V_+V_- \det M .}

For $N_f > N_c$, we may not be able to write a superpotential that
will reproduce the classical Higgs branch, and a different description
may be necessary. In any case, for $N_f \geq N_c$ we expect to remain
with a one dimensional Coulomb branch, parameterized by $Y$, which
intersects the Higgs branch (which is the same as the classical Higgs
branch, of dimension $2N_fN_c-(N_c^2-1)$) at $Y=0$. 
There are also mixed branches, in
which both $Y$ and some of the mesons acquire VEVs. For instance, for
$N_f = N_c$, even when $Y$ is non-zero $M$ can obtain a VEV as long as
its rank obeys ${\rm rank} (M) \leq N_c-2$, corresponding to the
classical condition for having (at least) an unbroken $U(1)$.
We expect the
origin of moduli space to correspond to some interacting SCFT for all
$N_f \geq N_c$. 

Let us now analyze in detail how this description comes about for the
case of $N_c=3$. In this case the Coulomb branch is divided into two
regions, according to the sign of $\phi_2$. Let us denote the
instanton factors \suninj\ by $Y_1$ and $Y_2$ for $\phi_2 > 0$, and by
$\tilde{Y}_1$ and $\tilde{Y}_2$ for $\phi_2 < 0$. As discussed above,
$Y = Y_1 Y_2 = \tilde{Y}_1 \tilde{Y}_2$ may be continuously defined
throughout the Coulomb branch. The global charges of these fields may
be determined by looking at the instanton zero modes, and are given by
(in the same conventions as above)
\eqn\globcharges{\matrix{& U(1)_R & U(1)_A \cr
\det M & 0 & 2N_f \cr
Y & 2N_f-4 & -2N_f \cr
Y_1 & -2 & 0 \cr
Y_2 & 2N_f-2 & -2N_f \cr
\tilde{Y}_1 & 2N_f-2 & -2N_f \cr
\tilde{Y}_2 & -2 & 0. \cr}}

Along the subspace $\phi_2=0$, we have a $U(1)$ theory with $N_f$
massless electrons, so we expect the perturbatively-corrected moduli
space in this region to be described by a superpotential of the form
$W = -N_f (V_+ V_- \det M )^{1/N_f}$, for some variables $V_\pm$
which semi-classically satisfy $V_+ \sim e^{\alpha \Phi_2 /
g^2}$ (for $\phi_2 \gg 0$) and $V_- \sim e^{-\alpha \Phi_2 / g^2}$
(for $\phi_2 \ll 0$) for some constant $\alpha$. 
For $N_f\geq 3$ there are also baryonic operators which can appear in
the superpotential; we set these to zero here.
In both regions of the moduli space, we can express
$e^{\Phi_2}$ (and, therefore, $V_\pm$) semi-classically in
terms of the instanton factors of this region, and determine
$\alpha=3/2$ by using the global symmetries. Solving these relations
for $Y_2$ and $\tilde{Y_1}$, and inserting the appropriate instanton
contributions, we find
\eqn\spotthree{W = -N_f  (V_+ V_- \det M )^{1/N_f} + \lambda_1(Y -
Y_1^2 V_+^2) + \lambda_2(Y - \tilde{Y}_2^2 V_-^2) + {1\over Y_1} +
{1\over \tilde{Y}_2},} where $\lambda_1$ and $\lambda_2$ are Lagrange
multipliers which enforce the relations between our variables, and
reduce the number of independent variables on the Coulomb branch to
three, as expected (generally each splitting of the Coulomb branch
adds one independent variable, so we expect to find $(N_c-1)+(N_c-2) =
2N_c-3$ independent variables parameterizing the Coulomb branch). It
is easy to see that this superpotential is compatible with adding mass
terms $W = \Tr ~m M$ and integrating out massive quarks to reduce
$N_f$, including the flow to the $N_f=0$ theory described above. It is
also compatible with giving one meson a VEV and flowing to the $SU(2)$
theory described in \S6.

\subsec{Theories with real mass terms for the quarks}

As discussed above, we can also add real mass terms for the quarks,
and we will discuss here only mass terms corresponding to background
vector-like fields, which do not break CP invariance. Again, it is
easy to determine semi-classically the number of quark zero modes for
each instanton, using the analysis of
\callias\ \foot{This may also easily be seen in the brane construction
of these theories.}.
A quark with real mass
$\tilde{m_f}$ will have zero modes in the background of an instanton
corresponding to a root $\phi_j-\phi_k$ if and only if $\phi_j >
\tilde{m}_f > \phi_k$ (and, in this case, we will have two zero modes,
one {}from the quark and one {}from the anti-quark chiral
multiplet). Whenever some $\phi_i=\tilde{m}_f$, we get a $U(1)$ theory
with a massless electron, and the Coulomb branch will split.

The analysis of these theories can be performed as described in the
previous section. We should first find a superpotential describing the
perturbatively-corrected moduli space, and then add to it the
instanton corrections to determine the full dynamics of the
theory. The construction of such theories is complicated in general,
and will not be presented here. However, as with the massless case,
the general form of the space of vacua of the theory can be determined
by simple arguments. Any region of the Coulomb branch where
semi-classically all masses are larger than $\phi_i$ or smaller than
$\phi_{i+1}$ will be lifted by the instanton corresponding to the root
$\phi_i-\phi_{i+1}$, which has no quark zero modes there. Unlike
the massless case, here higher dimensional subspaces of the Coulomb
branch may also remain unlifted. For example, for $N_f \geq N_c-1$
with all masses different, we expect to have $(N_c-1)$-dimensional
subspaces of the Coulomb branch which remain unlifted (though they may
still mix with the Higgs branches), corresponding (for instance) to
\eqn\unlifted{\phi_1 >
\tilde{m}_1 > \phi_2 >
\tilde{m}_2 > \cdots > \tilde{m}_{N_c-1} > \phi_{N_c}.}

\subsec{$d=4$ $N=1$ SQCD theories compactified on a circle}

The analysis of these theories is similar to the analysis of \S7.
For simplicity we will discuss here only the cases with no real
masses. Then, the effect of the finite radius is (for any $N_f$) just
to add to the superpotential we wrote in the previous sections
a term $W = \eta Y$ where $\eta =
e^{-1/Rg^2} \sim e^{-1/g_4^2} \sim \Lambda_4^{3N_c-N_f}$. As in \S7,
this term arises {}from instantons twisted by a non-trivial gauge
transformation on the circle \leeyi.

For $N_f=0$, as discussed in \kv, this term leads to the existence of
$N_c$ stable vacua for any value of $R$. Integrating out the massive
fields we find $W \sim \Lambda_4^3$, as expected. For $0 < N_f < N_c$,
adding this term leads to no stable vacua. We can integrate out all
the fields parameterizing the Coulomb branch, and find the expected
superpotential \ads\ $W \sim (\Lambda_4^{(3N_c-N_f)} / 
\det M)^{1/(N_c-N_f)}$.

For $N_f=N_c$, adding $W_\eta$ to \spotmid\ yields $W = Y (\eta -
(\det M - B \tilde{B}))$. In the $d=4$ limit, integrating out $Y$,
this leads to the known quantum constraint on the moduli space
\nati. For finite $R$, we find a similar constraint on the Higgs
branch, which becomes the classical constraint as $R \to 0$, and at
$R=0$ the theory grows an additional one dimensional Coulomb
branch. For $N_f = N_c+1$, we can similarly find (if we put in the
baryons in the correct way to reproduce the classical Higgs branch in
the $d=3$ theory) the $d=4$ superpotential $W = (\det M - B_i
M^i_{\tilde j} \tilde{B}^{\tilde j}) /
\Lambda_4^{2N_c-1}$ \nati.

For higher values of $N_f$, we have not been able to find a good
description of the $d=3$ theories, and also in $d=4$ we do not have a
description of the theory in terms of gauge invariant variables. In
$d=4$ there is a dual theory which flows to the same fixed point in
the IR \sem, but it is not clear if the same type of duality works in
$d=3$.

\bigskip
\centerline{{\bf Acknowledgments}}

We would like to thank T. Banks, S. Shenker, C. Vafa, and E. Weinberg
for useful discussions.  K.I. and N.S. would like to thank the
Harvard theory group for hospitality during the completion of this
paper.  The work of K.I. and M.J.S. were supported by National Science
Foundation grant NSF PHY-9513835 and by the WM Keck Foundation.  The
work of K.I. was also supported by an Alfred Sloan Foundation
Fellowship and the generosity of Martin and Helen Chooljian.  The work
of M.J.S. was also supported by NSF grant PHY-9218167.  The work of
O.A. and N.S. was supported in part by DOE grant DE-FG02-96ER40559.
The work of A.H. is supported in part by NSF Grant PHY-9513835.

\listrefs

\end